\documentclass[twocolumn,showpacs,amsmath,amssymb,pra,floatfix]{revtex4}

\usepackage{bm}
\usepackage{epsf}
\usepackage[english]{babel}
\usepackage{graphicx}
\def\op#1{\hat{\rm #1}}

\begin{document}


%
%

\title{Antiproton and proton collisions with the alkali metal atoms Li, Na,
  and K} 

\author{Armin L\"uhr}
\author{Alejandro Saenz}


\date{\today}



\pacs{34.50.Fa}

\begin{abstract}\label{txt:abstract}
Single-electron ionization and excitation cross sections as well as
cross sections for excitation into the first excited p state of the alkali
metal atoms Li(2s), Na(3s) and K(4s) colliding with antiprotons and protons
were calculated using a time-dependent channel-coupling approach. For
antiprotons an impact-energy range from 0.25 to 1000\,keV and 
for protons from 
2 to 1000\,keV was considered. The target atoms are treated as effective
one-electron systems using a model potential. 
The results are compared with theoretical and experimental data from
literature and calculated cross sections for antiproton-hydrogen
collisions. For proton collisions a good overall agreement is found which
confirms the present numerical approach, whereas discrepancies are found
between the present antiproton cross sections and those calculated by Stary et
al.,\, J.\,Phys.\,B {\bf 23}, 263 (1990).
\end{abstract}

\maketitle


\setcounter{table}{0}
\clearpage

\section{Introduction}
%
%
Collisions with alkali metal atoms as targets have been studied in numerous
experimental and theoretical works over many years. Among these studies a
number of efforts deal with proton--alkali metal atom collisions %
%
\cite{flair:shah85,flair:auma85,flair:auma84,flair:zapu05,flair:auma87,flair:giel91,flair:schw99,flair:wutt97,flair:taba88,flair:morg85,flair:bran98,flair:nagy00,flair:shin87,flair:jain95,flair:shin86,flair:frit84}  
 and a smaller number of attempts address collisions including antiprotons as
 projectiles %
\cite{flair:star90,flair:mcca93}
. One reason for the  attractivity of alkali metal atoms is that they are
relatively 
easy to access experimentally as well 
as theoretically which opens up the possibility for detailed comparisons. The
given shell structure of the alkali metals suggests in a theoretical
description the 
application of a quasi-one-electron model for the outermost loosely bound
electron. The electron is then described by means of a model potential
formed by 
the Coulomb potential of the nucleus and an effective potential representing
the frozen inner-shell electrons. In particular, Li and Na atoms 
colliding with protons and electrons have been in the focus of the
investigations so far. Whereas the literature on alkali metal atoms collisions
dealing with antiprotons as projectile is still sparse compared to the
treatment of protons and electrons. In 
order to obtain cross sections for ionization of Li by antiproton impact a
continuum-distorted-wave eikonal-initial-state model has been used by
McCartney et al.\ \cite{flair:mcca93}. Furthermore, an optical-potential
description of  collisions of antiprotons with Li and Na has been provided by 
Stary et al.\ \cite{flair:star90}. No experimental data are available for the
considered antiproton--alkali metal atom collision systems yet due to the lack
of appropriate low-energy antiproton sources. This may also
be the reason for the relatively small interest in antiproton--alkali metal
collisions compared to their proton counterparts until now. However, 
the upcoming Facility for Antiproton and Ion Research (FAIR) with its
incorporated Facility for Low energy Antiproton and Ion Research (FLAIR)
\cite{flair:flai} 
will provide the necessary experimental conditions in the near future and is
therefore expected to attract further attention to the field of antiproton
collisions.   

The primary motivation of this work is to shed more light on the
antiproton--alkali metal collision systems and to provide a consistent data
base for Li(2s), Na(3s) 
and K(4s) atom collisions with antiprotons and protons in a large energy range.
It starts at low energies $E=0.25$\,keV where the collision processes depend
considerably on the projectile and ranges up to
high energies $E=1000$\,keV where the antiproton and proton collision systems
are supposed to show the same behavior due to the expected applicability of
the first Born approximation. The calculations for collisions with proton
projectiles are considered to be valuable in two aspects. On the one hand, the
proton results -- especially for Li targets -- can be compared in detail
with literature values. This way the proton results can be utilized in order to
test the present method and its implementation which is the same for protons and
antiprotons. Furthermore, new theoretical ionization and excitation cross
sections for proton collisions -- especially for K targets -- are provided
which to the authors' knowledge  were not fully known in the energy range
considered here.  

Besides the obvious 
similarities of protons and antiprotons as projectiles they differ mostly in
their capture behavior. First, only antiprotons can annihilate with protons
of the atomic nucleus. Since it is known that the process of annihilation is
only likely to occur at very low energies \cite{flair:cohe04} it is not
included in this investigation.  
Second, in the case of proton collisions electron capture by the projectile
from the target atom is possible. This process plays a dominant role for
low-energy collisions. Hence, a two-center approach appears to be most
promising for low-energy proton collisions. 
However, at low energies the present calculations concentrate on 
antiproton collisions only. Therefore, a basis expansion which is
centered solely on the target alkali metal atom for both antiproton and proton
projectiles is used. Thereby, limitations pertinent to a molecular approach at
high energies are avoided. Furthermore, the same method can be used for
antiproton and proton collisions which may be confirmed by a detailed
comparison of present proton results with literature data. A detailed analysis
of the electron capture process for proton scattering, however, lies beyond
the scope of this work.   

The paper is organized as follows: Sec. \ref{sec:method} explains how
the alkali metal atoms are described and reports on the computational
approach. Sec. \ref{sec:results} considers the convergence behavior of the
present antiproton and proton results. Subsequently, the 
calculated cross sections are presented and compared to literature
data. Finally, the present results for antiproton--alkali metal collisions are 
discussed and a comparison with a hydrogen atom as target is made. Sec.
\ref{sec:summary}  concludes on the present findings.  Atomic units are used
unless it is otherwise stated.

%
\section{Method}
\label{sec:method}
%
%
In this work the target atoms are treated as effective one-electron atoms.
The valence electron is exposed to a model potential $V_{\rm mod}$ suitable
for alkali metal atoms which describes its interaction with the nucleus as
well as 
with the remaining core electrons. Additionally, core polarization effects are
included. The employed model potential was proposed by 
Klapisch \cite{aies:klap71}. The used potential parameters are
given in \cite{aies:magn99a}.  The effect of the spin-orbit coupling is not
included in the present approach. The energies of the states with principal
quantum number $n<6$ for the alkali metal atoms Li, Na and K which were
achieved with this approach are given in table \ref{tb:binding-energies}
together with compiled values of the NIST data bank \cite{flair:nist}. In the
case of energy level splitting due to spin-orbit coupling the present energies
are compared to the lower lying reference energies. The largest relative energy
splittings of the reference data due to the spin-orbit coupling of the
energies given in table
\ref{tb:binding-energies} are 0.002\,\%, 0.1\,\% and 0.4\,\% for the
energetically lowest lying p states of Li, Na and K, respectively.
Particularly for Li there is a very good agreement with the data given by
NIST. But also for the other two atoms the deviation from the literature
values remains at maximum around one per cent.
%
\begin{table}[t]
  \centering
  \begin{tabular}{c@{\hspace{0.3cm}}|@{\hspace{0.2cm}}l@{\hspace{0.2cm}}|@{\hspace{0.2cm}}l@{\hspace{0.2cm}}|@{\hspace{0.2cm}}l}
    \hline
    \hline
    $n\,l$ & Li calc.&Na calc. &K calc. \\
           & Li ref.\ \cite{flair:nist} &Na ref.\  \cite{flair:nist}&K ref.\  \cite{flair:nist}\\
    \hline
    2s  &-0.198477&&\\
        &-0.198142&&\\
    2p  &-0.130482&&\\ 
        &-0.130236&&\\ 
    3s  &-0.074362&-0.189163&\\ 
        &-0.074182&-0.188858&\\
    3p  &-0.057364&-0.111760&\\  
        &-0.057236&-0.111600&\\
    3d  &-0.055605&-0.056071&-0.061596\\ 
        &-0.055606&-0.055937&-0.061397\\ 
    4s  &-0.038694&-0.071754&-0.160105\\ 
        &-0.038615&-0.071578&-0.159516\\
    4p  &-0.032036&-0.051075&-0.100434\\ 
        &-0.031975&-0.050951&-0.100352\\ 
    4d  &-0.031274&-0.031531&-0.034954\\ 
        &-0.031274&-0.031442&-0.034686\\ 
    4f  &-0.031254&-0.031267&-0.031337\\ 
        &-0.031243&-0.031268&-0.031357\\ 
    5s  &-0.023677&-0.037656&-0.064121\\ 
        &-0.023637&-0.037584&-0.063713\\
    5p  &-0.020408&-0.029265&-0.047187\\ 
        &-0.020374&-0.029202&-0.046969\\
    5d  &-0.020013&-0.020160&-0.022158\\ 
        &-0.020013&-0.020106&-0.021983\\ 
    5f  &-0.020002&-0.020010&-0.020048\\ 
        &-0.019969&-0.020011&-0.020062\\
    \hline
    \hline
  \end{tabular}
  \caption{Calculated binding energies (Hartree) for Li, Na and K using a %
    Klapisch-model potential. The reference data is taken
    from the NIST data tables \cite{flair:nist}. In the case of energy level
    splitting due to spin-orbit coupling only the energetically lower lying
    reference energy is given.\label{tb:binding-energies}} 
\end{table}

In order to describe the collision process the relative motion of the heavy
particles is approximated by a classically trajectory (CTA) also
referred to as the impact-parameter representation. The projectile is assumed
to move on a classical rectilinear 
trajectory with a constant velocity ${\bf v}$ parallel to the $z$ axis. The
internuclear distance vector $\mathbf{ R}$ is given by ${\mathbf{ R=b+v}}t$,
where $\mathbf{b}$ is the impact-parameter vector along the $x$ axis and $t$
the time.  

The time-dependent Schr\"odinger equation 
\begin{equation}
  \label{eq:tdSE}
  i \, \frac{\partial}{\partial t} \, \Psi({\bf r},t) = 
  \left (\, \op{H}_0 + \op{V}_{\rm int}({\mathbf{ r,R}}(t))\, \right )\,
  \Psi({\bf r},t)\,
\end{equation}
of the target atom interacting with the projectile is solved. The atomic
Hamiltonian of the target atom is defined as 
\begin{equation}
  \label{eq:atomic_Hamiltonian}
 \op{ H}_0 = -\frac{1}{2} \nabla^2 + \op{V}_{\rm mod}  
\end{equation}
and the time-dependent interaction between the projectile with the charge
$Z_p$ and the target atom as
\begin{equation}
  \label{eq:interaction_potential}
  V_{\rm int}({\mathbf{ r,R}}(t)) = 
           \frac{-Z_p}{\left| \mathbf{r - R}(t) \right|} + 
           \frac{Z_p}{|\mathbf{R}(t)|}
\end{equation}
where ${\bf r}$ is the spacial coordinate of the explicitly treated valence
electron.

The total wavefunction $\Psi({\bf r},t)$ is expanded as 
\begin{equation}
  \label{eq:total_wavefunction}
 \Psi({\bf r},t) = \sum_{nlm}\, c_{nlm}(t)\,
                   \phi_{nlm}(\mathbf{r})\,
                   \exp(-i\,\epsilon_{nl}\,t)
\end{equation}
using the expansion coefficients $c_{nlm}(t)$.
Here, the $\phi_{nlm}$ are eigenstates of the Hamiltonian $\op{H}_0$ with the
energy $\epsilon_{nl}$ obtained with the model potential. The $n,\,l,\,m$ are
the principal one-electron quantum number, angular momentum, and its projection
on the $z$ axis, respectively. The $\phi_{nlm}$ can be further expanded as 
\begin{equation}
  \label{eq:atomic_wavefunction}
  \phi_{nlm}(\mathbf{r}) = g_{nl}(r) \, \frac{1}{\sqrt{2(1+\delta_{m,0})}} \,
                          \left[ \, 
                            (-1)^m Y_l^m(\Omega) + Y_l^m(\Omega)  \,
                          \right], 
\end{equation}
where the $Y_l^m(\Omega)$ are the spherical harmonics depending on the
angular part $\Omega$ of $\bf r$. In Eq.\ (\ref{eq:atomic_wavefunction}) the
symmetry of the Hamiltonian under reflection at the collision plane given by
$\bf b$ and $\bf v$ is used in order to reduce the number of states. The
radial function $g_{nl}(r)$  is further expanded in terms of ($k-1$)th order
$B$-spline functions. Converged results were found confining the entire space
of the electron to a sphere of radius $r_{\rm max}=200$ with fixed boundary
conditions. The range $[0,r_{\rm max}]$ is divided into $N_r - 1$ intervals
between the knot points ($r_1=0,\ldots,r_{N_r}=r_{\rm max}$). 
%
\begin{equation}
  \label{eq:radial_function}
  g_{nl}(r) = \sum_j^{k+N_{r}-2} \,a^j_{nl}\, \frac{B^k_j(r)}{r}                
\end{equation}
The expansion coefficients $a_{nl}^j$ are determined by diagonalizing the
atomic Hamiltonian $\op{H}_0$.

Substitution of the total wavefunction $\Psi$ in Eq.\ (\ref{eq:tdSE}) by its
expansion given in Eq.\ (\ref{eq:atomic_wavefunction}) results in a system of
coupled equations,
\begin{eqnarray}
  \label{eq:coupled_equaitons}
  i\,\frac{d}{d t}c_{n'l'm'}(t) &= &
           \sum_{nlm}  c_{nlm}(t) \,
           \langle\,\phi_{n'l'm'}\,| \,\op{V}_{\rm int}\,
           |\,\phi_{nlm}\,\rangle\,\\ 
    &  &\quad \times \ \exp[\,i(\epsilon_{n'l'} -
    \epsilon_{nl})\,t\,]\,.\nonumber  
\end{eqnarray}
These coupled equations are solved for $N_b$ fixed values of the impact
parameter $b$ with the 
initial condition $c_{nlm}(t=-\infty,b)=\delta\,_{nlm,\,n_il_im_i}$ where the
index $n_il_im_i$ represents the initial state of the atom. The transition
probability $P_{nlm}(b)$ into the atomic state $(n,l,m)$ after the collision
is given by 
\begin{equation}
  \label{eq:transition_probability}
  P_{nlm}(b) = |\,c_{nlm}(t=+\infty,b) \,|^2\,.
\end{equation}
The cross section $\sigma_{nlm}$ for the transition into the state $(nlm)$
follows from 
\begin{equation}
  \label{eq:state_cs}
  \sigma_{nlm} = 2\, \pi\, \int {\rm d}b\,b\,P_{nlm}(b)\,.
\end{equation}
The total cross sections for ionization
\begin{equation}
  \label{eq:ionization_cs}
  \sigma_{\rm ion} = \sum_{\epsilon_{nlm}>0} \sigma_{nlm}
\end{equation}
and for excitation of the target atom
\begin{equation}
  \label{eq:excitation_cs}
  \sigma_{\rm ex} = \sum_{\epsilon_{n_il_i}<\epsilon_{nlm}<0} \sigma_{nlm}
\end{equation}
can readily be calculated where $\epsilon_{n_il_i} $ is the energy of the
initial state of the target atom.

\section{Results and discussion}
\label{sec:results}
In what follows, first the convergence behavior and second
the dependence of the ionization and excitation cross sections on the
impact parameter $b$ for different impact energies is
investigated. Thereafter, the results for proton and antiproton collisions
with the alkali metal target atoms Li, Na and K are presented and the findings
are compared with data from literature. Additionally, the antiproton cross
sections are compared with calculations including a hydrogen atom as target.

\subsection{Convergence behavior and b-dependent transition probabilities} 
\label{subsec:convergence-b-dependence}
\begin{figure}[t]
    \begin{center}
      \includegraphics[width=0.49\textwidth]{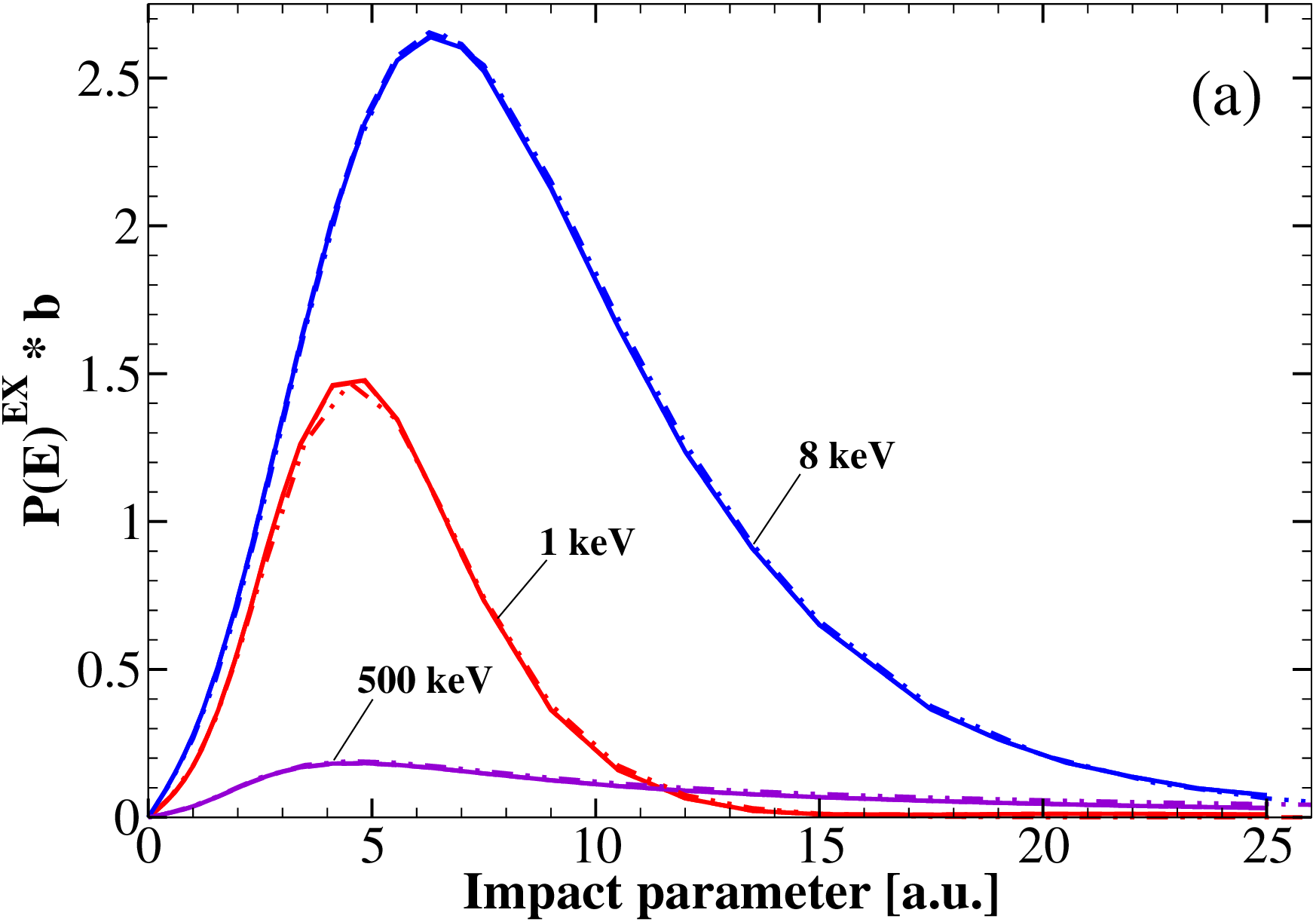}
      \includegraphics[width=0.49\textwidth]{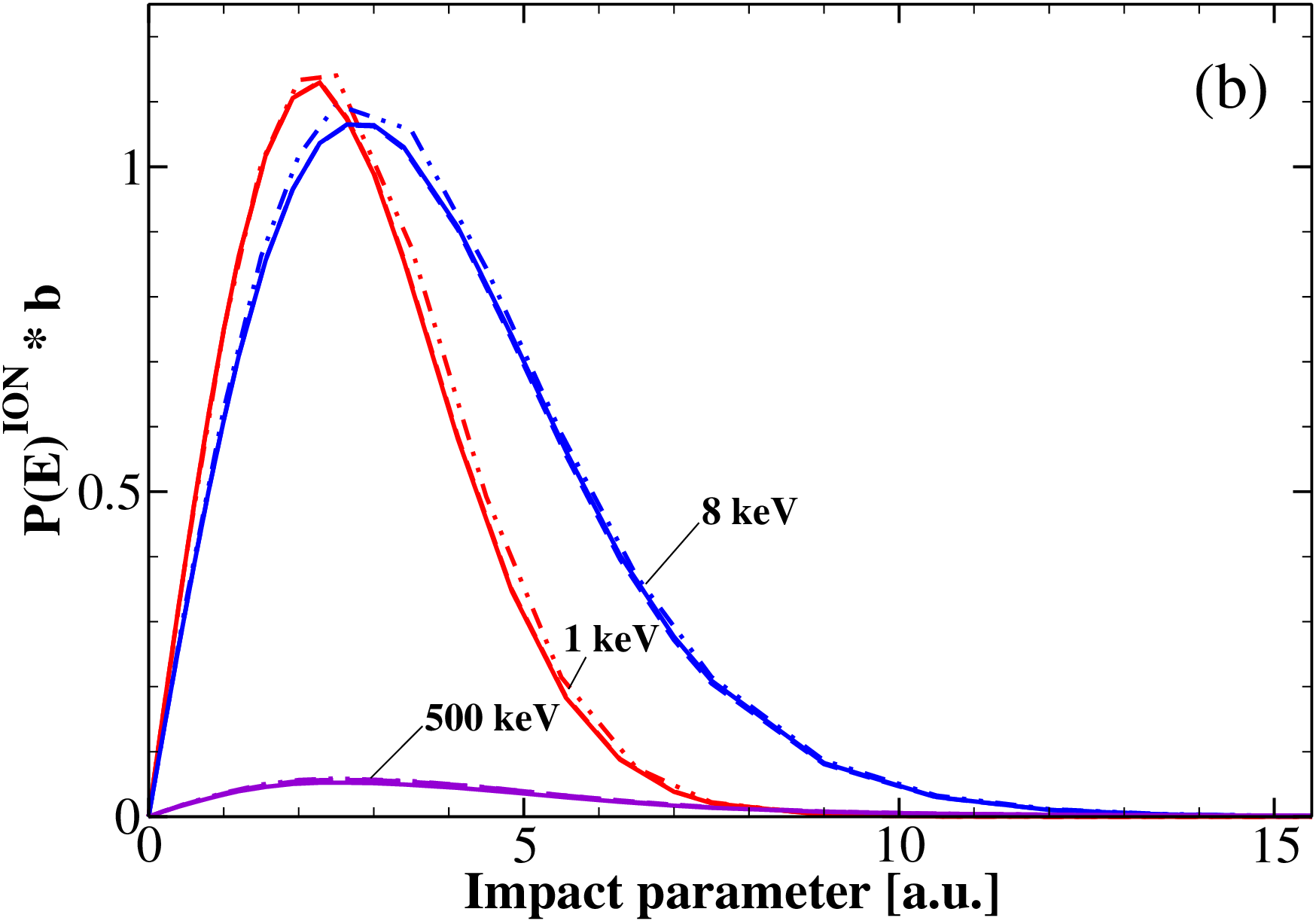} 
      \caption{(Color online) \=p - Li(2s) collision: The
        convergence behavior at different impact energies is 
        shown for the three basis sets A4 (--), A6 ({\bf- -}) and A8
        ({\bf$\cdot-\cdot$}) with maximum angular momenta $l_{\rm max}=4$, 
        6 and 8, respectively. (a) Total excitation
        probability $P(E)^{\rm EX}$ 
        weighted with the impact parameter $b$ as a function of $b$.
        (b) As (a), but ionization probability $P(E)^{\rm ION}$.
        \label{fig:pb_b_lmax} }
    \end{center}
\end{figure}
\begin{figure}[t]
    \begin{center}
      \includegraphics[width=0.49\textwidth]{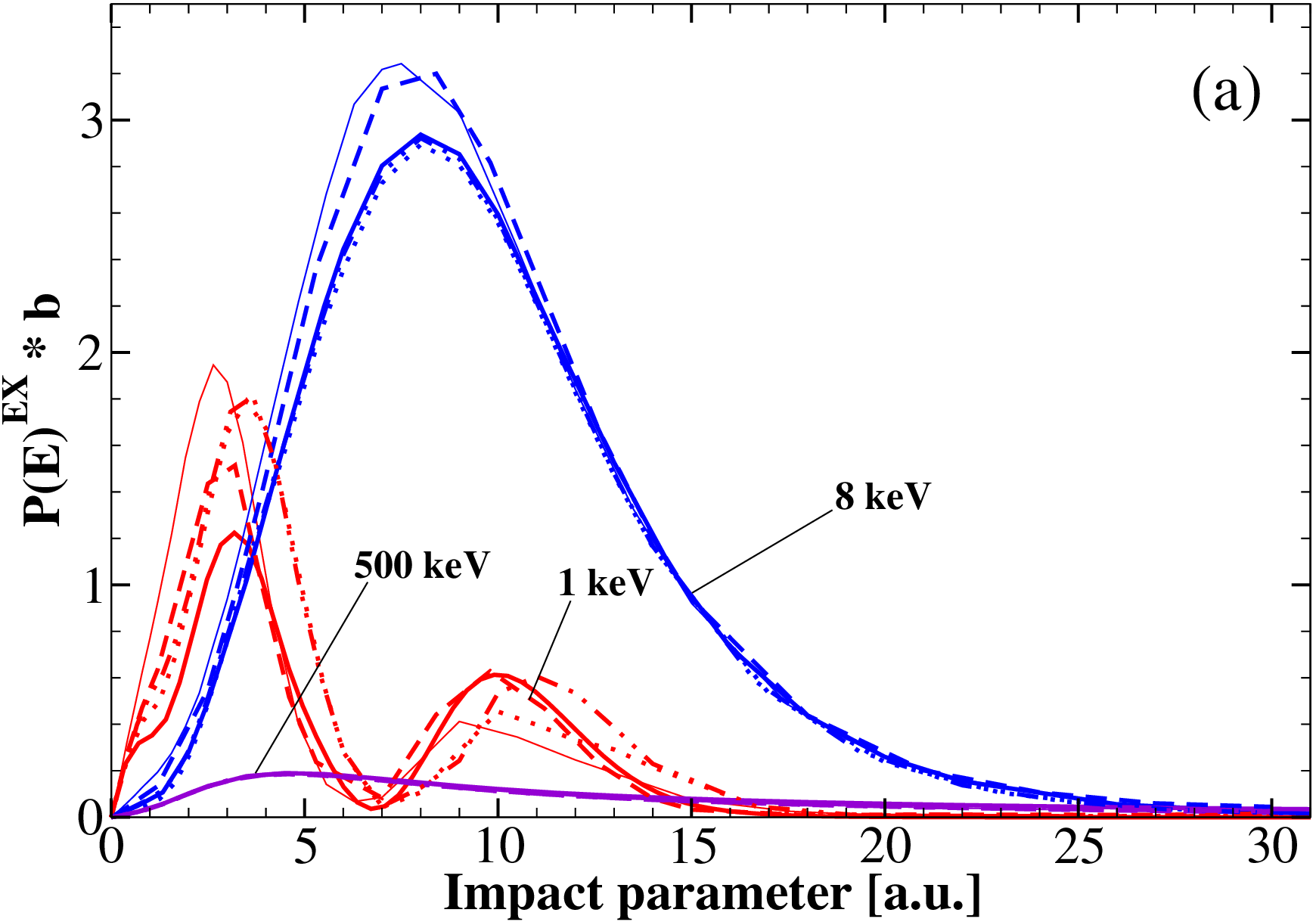}
      \includegraphics[width=0.49\textwidth]{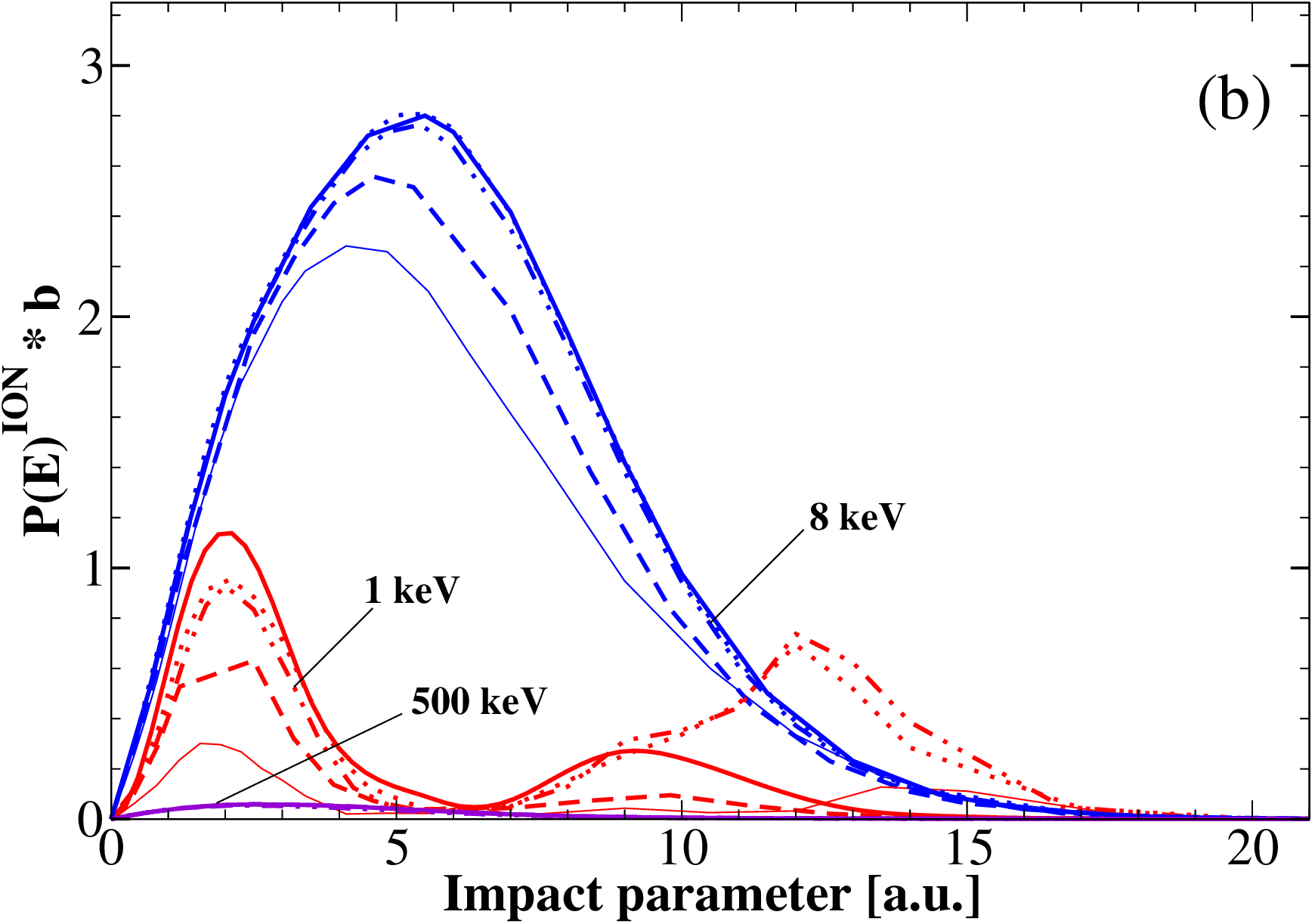} 
      \caption{(Color online) p - Li(2s) collision:  The
        convergence behavior at different impact 
        energies is shown for the five different basis sets P6 (-),
        P8 ({\bf- -}), P10a ({\bf$\cdot-\cdot$}), P14 ({\bf$\cdots$})  and P10b
        ({\bf-}) with maximum angular momenta $l_{\rm max}=6$, 8, 10, 14 and
        10
        , respectively. (a) Total excitation probability $P(E)^{\rm EX}$    
        weighted with the impact parameter $b$ as a function of $b$.  (b) As
        in (a), but ionization probability $P(E)^{\rm ION}$.
       \label{fig:pb_b_lmax_p}}
    \end{center}
\end{figure}
\begin{table}[b]
  \centering
  \begin{tabular}{lccc@{\hspace{0.2cm}}|@{\hspace{0.2cm}}lccc}  
    \hline
    \hline
    basis &$l_{\rm max}$ &$m_{\rm max}$&states & basis &$l_{\rm max}$ &$m_{\rm
      max}$&states \\ 
    \hline
    \,A4  &4  &4 & 810 &\,P6  &6  &6 & 1188  \\
    \,A6  &6  &6 & 1512&\,P8  &8  &8 & 1620 \\
    \,A8a &8  &8 & 2430&\,P10a&10 &3 & 2052 \\
    \,A8b &8  &3 & 1620&\,P14 &14 &3 & 2916 \\
          &   &  &     &\,P10b&10 &6 & 3024 \\
    \hline
    \hline
  \end{tabular}
  \caption{Parameters of the basis sets used for the
    convergence studies. Basis sets beginning with A (P) are used in
    calculation with antiprotons (protons). For each basis set the maximum
    angular quantum number $l_{\rm max}$, the maximum magnetic quantum
    number $m_{\rm max}$ and the total number of basis states are given. 
    \label{tb:basis-parameter}} 
\end{table}
In order to discuss the convergence of the results the behavior of the
product $P(b,E)\,b$ is investigated. This quantity results after integration
over $b$ -- 
cf.\ Eq.\ (\ref{eq:state_cs}) -- in the final cross section. Thereby, the
probability $P$ for a certain transition depends on the 
impact parameter $b$ and the impact energy $E$. In figures \ref{fig:pb_b_lmax}a
and \ref{fig:pb_b_lmax}b
the transition probabilities for total excitation and ionization, respectively,
of \=p - Li(2s) collisions calculated for three different energies are
presented. The results for p - Li(2s) collisions for the same parameters are
shown in figure \ref{fig:pb_b_lmax_p}.

In the case of antiproton collisions with lithium calculations for three
different basis sets A4, A6 and A8 with maximum angular momenta $l_{\rm max}=
4$, 6 and 8, respectively, are shown in figure \ref{fig:pb_b_lmax}. In table 
\ref{tb:basis-parameter} the maximum angular and magnetic quantum numbers as
well as the total number of basis states are given for the basis
sets. The  calculations using A4 and A6 are fully
converged. Therefore,  the range of integration has been changed from $-z_{\rm
  min}=z_{\rm max}=60$ 
to $-z_{\rm  min}=z_{\rm  max}=180$ for the calculation using the basis set
A8a. This results in slightly higher ionization and lower excitation
probabilities at low energies. It was found that the results for the
calculation using $l_{\rm max}=8$
converge quickly with increasing maximum projection of the angular momentum
$m_{\rm  max}$. Consequently, for the subsequent calculations dealing with
antiprotons as projectile the basis set A8b with $l_{\rm max}=8$, $m_{\rm
  max}=3$ and $z_{\rm max}=180$ was chosen resulting in a set of 1620 basis
functions.  

In the case of proton collisions with lithium calculations for five different
basis sets P6, P8, P10a, P14 and P10b different basis sets with maximum
angular momenta $l_{\rm max}= 6$, 8, 10, 14 and 10, respectively, are shown in
figure \ref{fig:pb_b_lmax_p}. The further parameters of these basis sets are
given again in table \ref{tb:basis-parameter}. In contrast to the antiproton
calculations, much higher  angular momenta are needed to 
achieve convergence -- especially for low energies $E\le 4$\,keV. The
results for the basis sets P10a and P14 both with $m_{\rm max}=3$ seem to be
converged with respect to $l$. Hence, $l_{\rm max}= 10$ was chosen but $m$ was
increased to $m_{\rm max}=6$ leading to the basis set P10b. Additionally, the 
integration range $z_{\rm max}$ was also enlarged to $z_{\rm max}=180$.  For all
considered energies with  $E>4$\,keV these parameters lead to satisfyingly
converged  results and were therefore used for all proton collision
calculations.   

From the previous analysis it can be concluded that convergence is achieved
much faster (i) for antiprotons than for protons, faster (ii) for excitation
than for ionization and (iii) at higher than at lower energies. 
Figures \ref{fig:pb_b_lmax} and
\ref{fig:pb_b_lmax_p} also provide insight into the physics of the
collision process. For high energies the same behavior for antiproton and
proton collisions can be observed. For energies below the validity regime of
the first Born approximation the transitions in antiproton collisions take 
place at smaller impact parameters compared to protons. For close encounters
which are more important for low energies the advent of the projectile inside
the orbit of the target electrons creates in the case of protons an increased
or for antiprotons a decreased binding of the electrons. This situation leads
to a decrease (p) or increase (\=p) of $P$ for small $b$ \cite{flair:knud92}
and a shift of the proton $P$ curves to larger $b$. The
ionization  probability is for p and \=p more concentrated in the vicinity of
the nucleus. This can be explained using the simple picture that the mean
velocity of the electrons close to the nucleus is higher than at larger
distances and therefore less energy transfer is required. On the other hand
 the excitation probability has a longer tail for large $b$ compared to
 ionization. Particularly at high energies care has to be taken that the 
calculations converge in the considered impact parameter range. 



\subsection{Cross sections for proton collisions }
\begin{figure}[t]
  \begin{center}
    \includegraphics[width=0.49\textwidth]{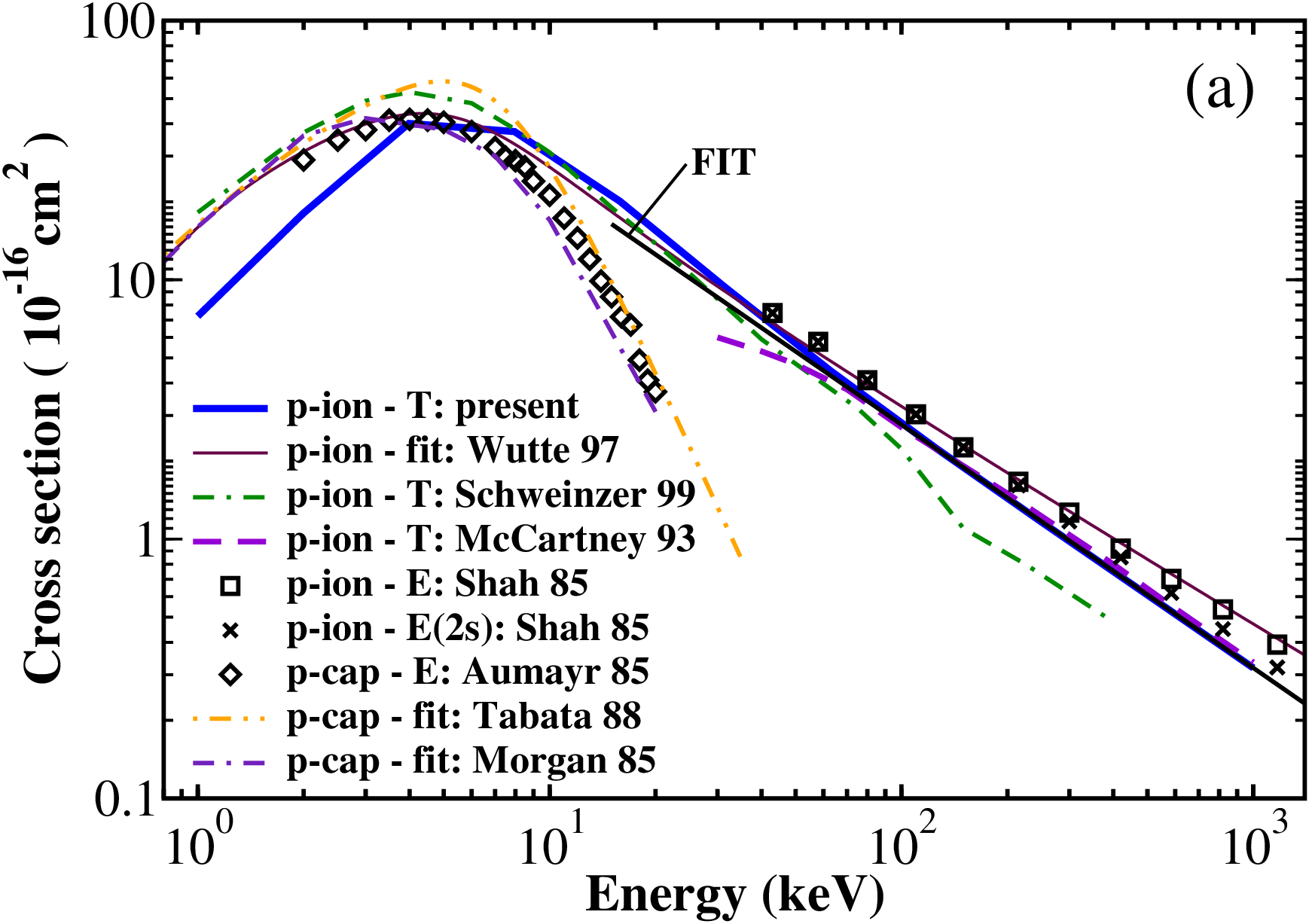}
    \includegraphics[width=0.49\textwidth]{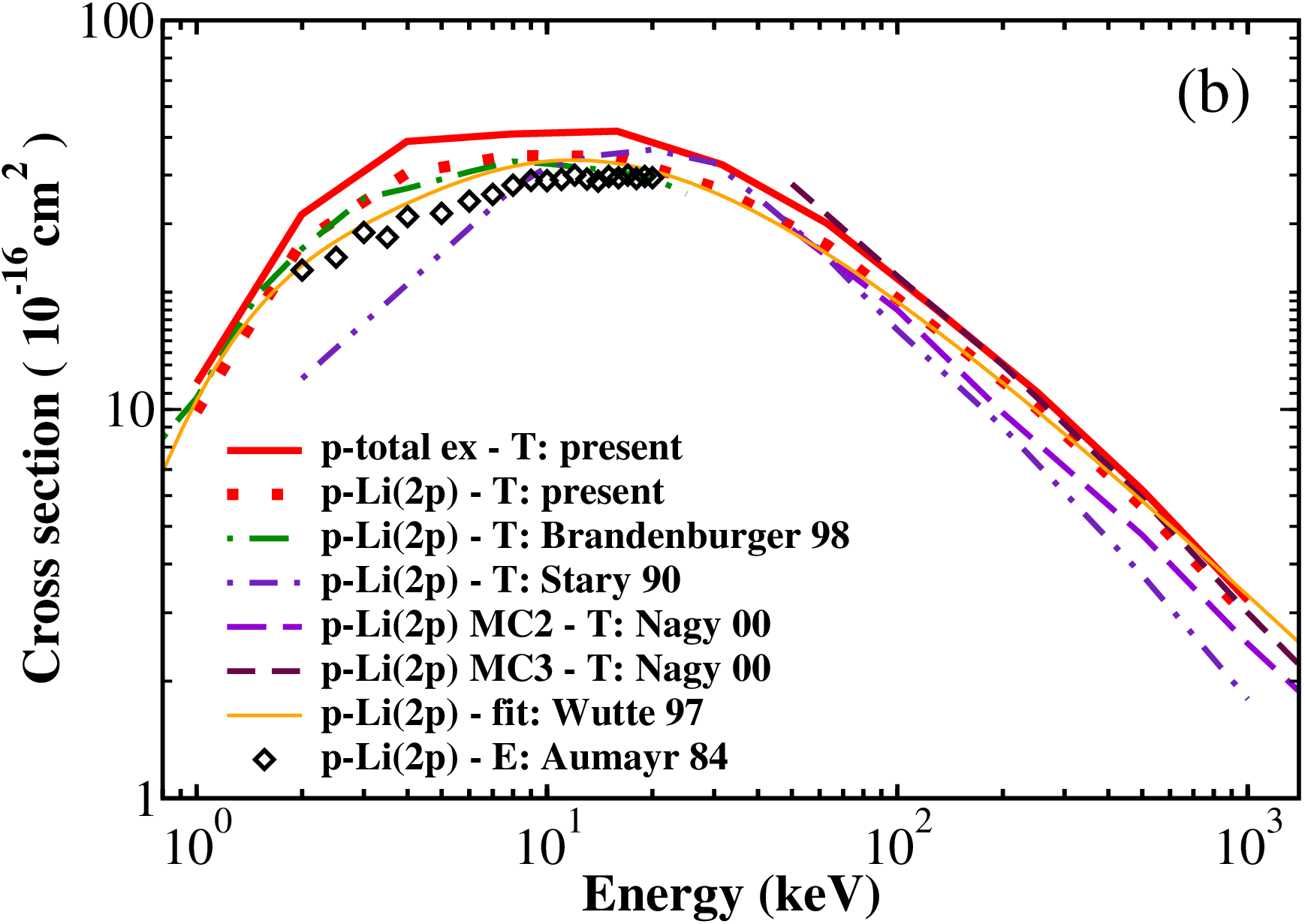}
    \caption{ (Color online) p - Li(2s): (a) Ionization and 
      capture. Theory(ionization): solid curve, present results; doubly-dashed
      dotted curve, Schweinzer et al.\ \cite{flair:schw99}; dashed 
      curve, McCartney et al.\ \cite{flair:mcca93}  (Li(2s)). Fit(ionization):
      thin solid 
      curve, Wutte et al.\ \cite{flair:wutt97}. Experiment(ionization):
      squares, Shah et al.\  \cite{flair:shah85}; crosses, Shah et al.\
      \cite{flair:shah85} (Li(2s)). Fit(capture): dashed
      doubly-dotted curve, Tabata et al.\ \cite{flair:taba88}; dashed dotted
      curve, Morgan et al.\  \cite{flair:morg85}. Experiment(capture):
      diamonds, Aumayr et al.\  \cite{flair:auma85}.
      (b) Total excitation and excitation into Li(2p). Theory(total
      excitation): solid curve, present results. Theory(excitation into
      Li(2p)): dotted curve, present results; dashed dotted curve,
      Brandenburger et al.\ \cite{flair:bran98}; dashed doubly-dotted curve,
      Stary et al.\ \cite{flair:star90}; long-dashed curve, MC2 Nagy et al.\
      \cite{flair:nagy00}; short-dashed curve, MC3 Nagy et al.\
      \cite{flair:nagy00}. Fit(Li(2p)): thin solid curve, Wutte
      et al.\  \cite{flair:wutt97}. Experiment(Li(2p)): diamonds, Aumayr et
      al.\  \cite{flair:auma84}.  \label{fig:cs_Li_l10_p} }  
  \end{center}
\end{figure}%
\begin{figure}[t]
  \begin{center}
    \includegraphics[width=0.49\textwidth]{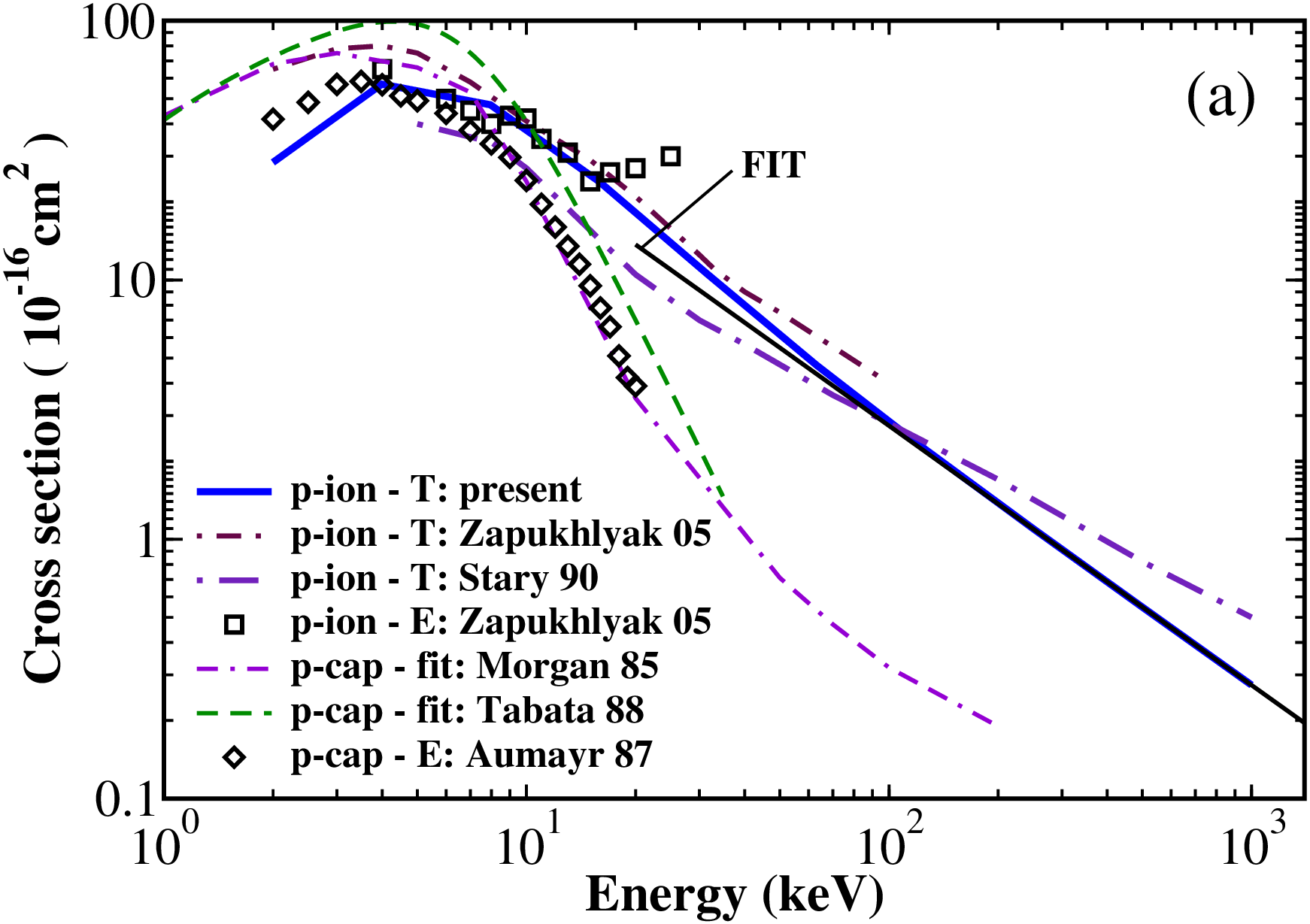}
    \includegraphics[width=0.49\textwidth]{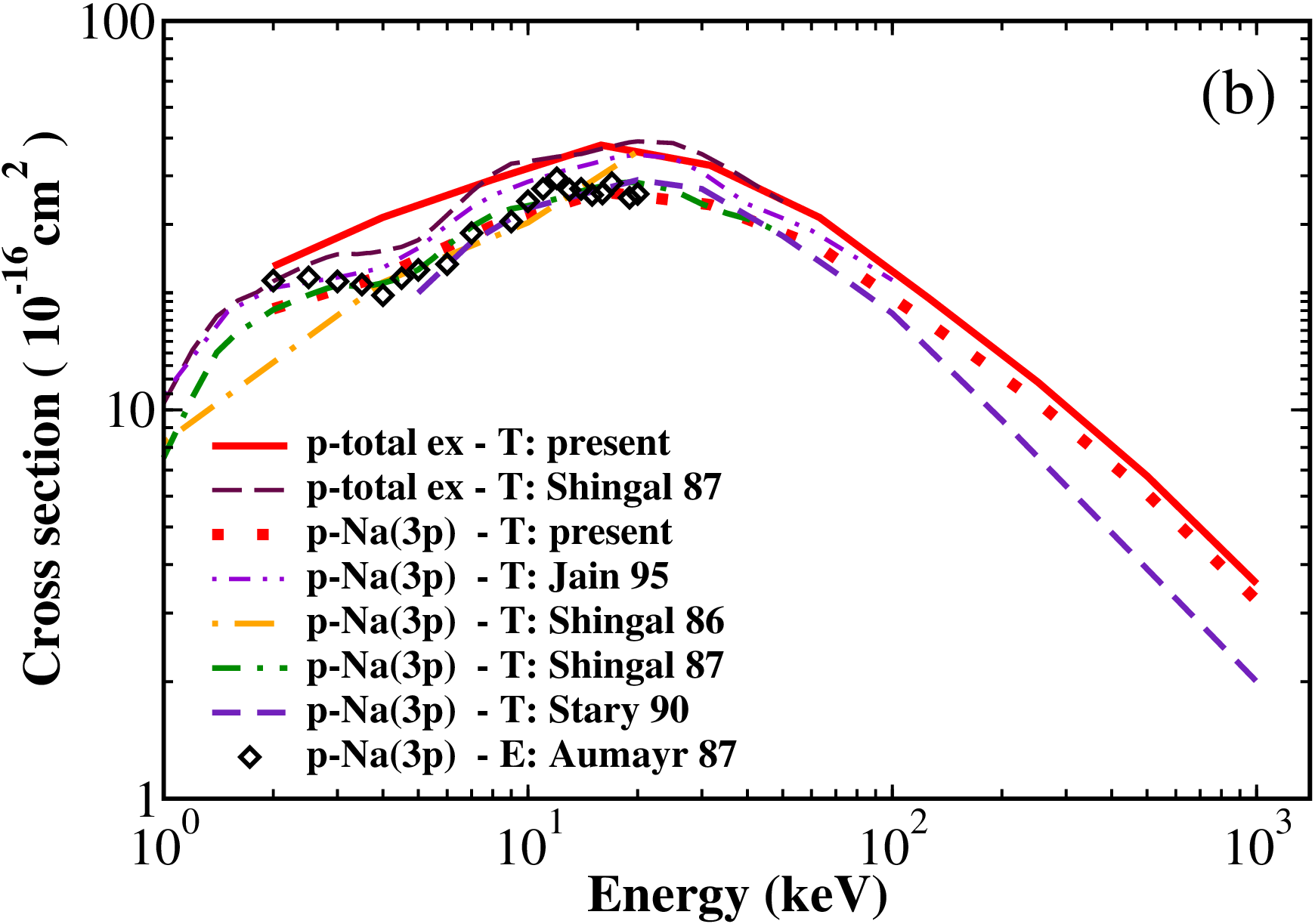}
    \caption{(Color online) p - Na(3s): (a) Ionization and
      capture. Theory(ionization): solid curve, present results; long-dashed
      dotted curve, Stary et al.\ \cite{flair:star90}; short-dashed dotted
      curve, Zapukhlyak et al.\ \cite{flair:zapu05}. Experiment(ionization):
      squares, Zapukhlyak et al.\ \cite{flair:zapu05}. Fit(capture): dashed
      curve, Tabata et al.\ \cite{flair:taba88}; doubly-dashed dotted curve,
      Morgan et al.\ \cite{flair:morg85}. Experiment(capture): diamonds,
      Aumayr et   al.\ \cite{flair:auma87}.
      (b) Total excitation and excitation into Na(3p). Theory(total
      excitation): solid curve, present results; long-dashed curve, Shingal et
      al.\ \cite{flair:shin87}. Theory(excitation into Na(3p)): dotted curve,
      present results; dashed doubly-dotted curve, Jain et al.\
      \cite{flair:jain95}; dashed dotted curve, Shingal et al.\  
      \cite{flair:shin86}; doubly-dashed dotted curve, Shingal et al.\
      \cite{flair:shin87}; short-dashed curve, Stary et al.\  
      \cite{flair:star90}. Experiment(Na(3p)): diamonds, Aumayr et al.\
      \cite{flair:auma87}. 
      \label{fig:cs_Na_l10_p} }  
  \end{center}
\end{figure}%
\begin{figure}[t]
  \begin{center}
    \includegraphics[width=0.49\textwidth,angle=0]{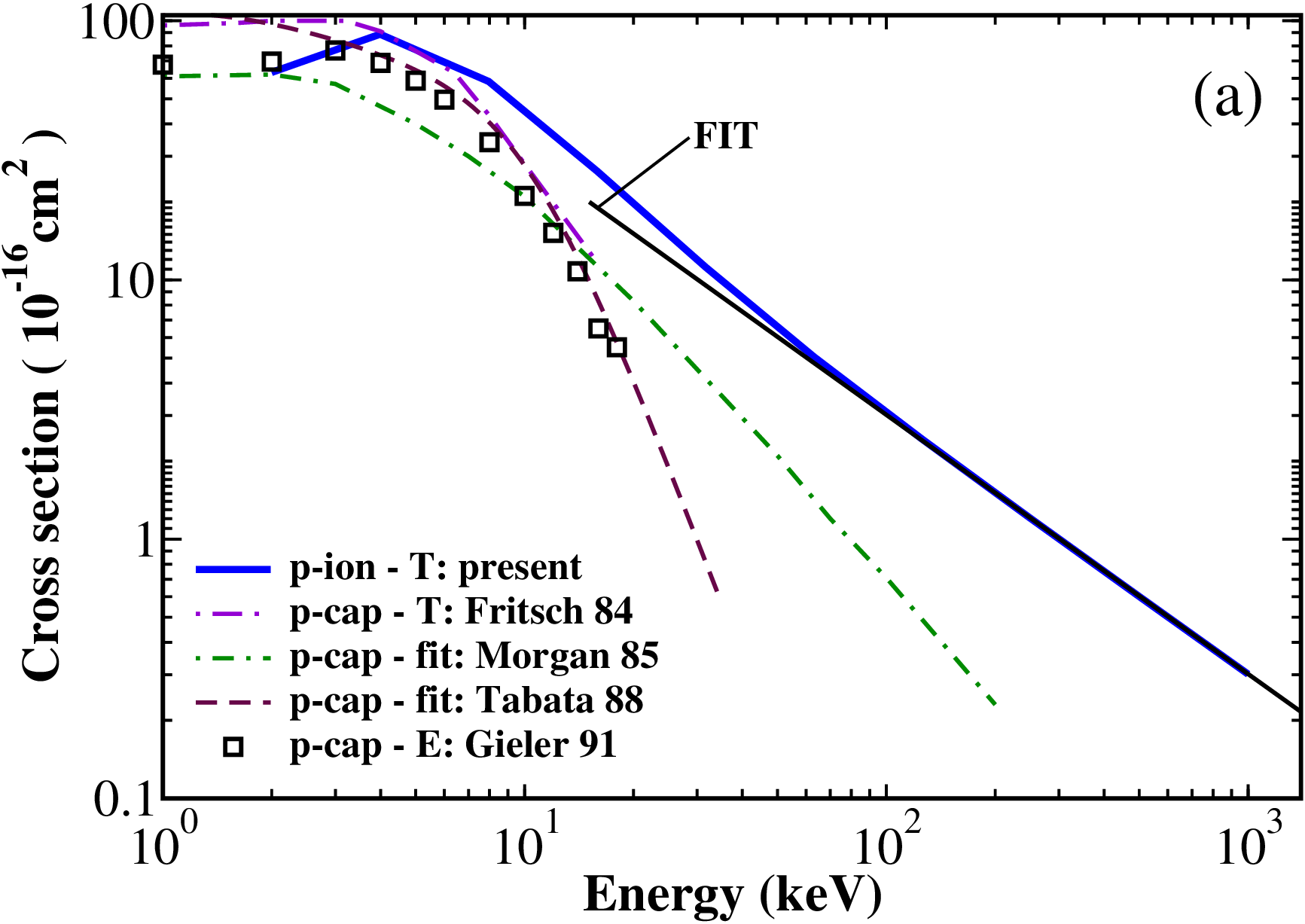}
    \includegraphics[width=0.49\textwidth,angle=0]{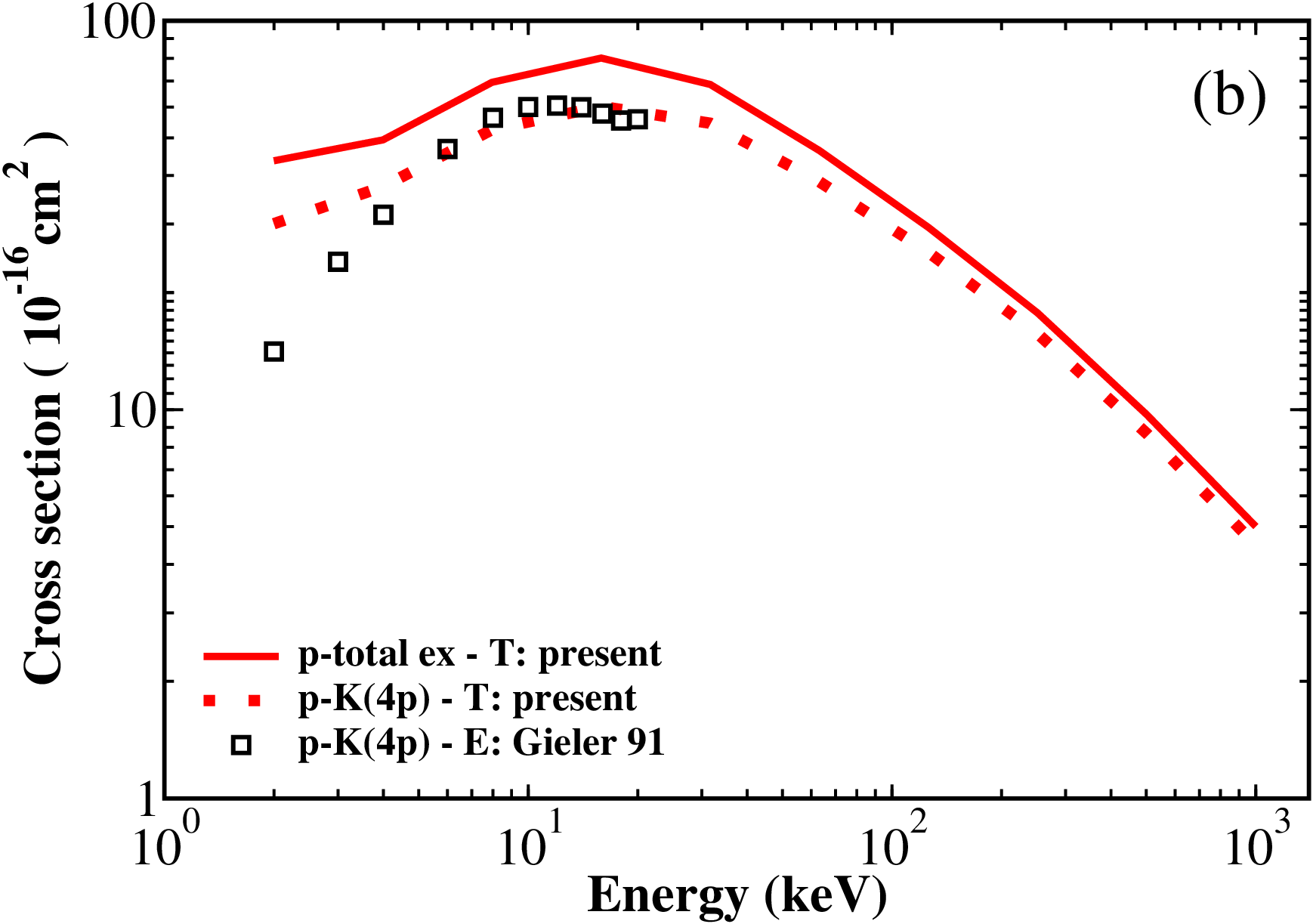}
    \caption{(Color online)  p - K(4s): (a) Ionization and
      capture. Theory(ionization): solid curve, present
      results. Theory(capture): long-dashed dotted curve, Fritsch
      \cite{flair:frit84}. Fit(capture): 
      dashed curve, Tabata et al.\ \cite{flair:taba88}; dashed dotted curve,
      Morgan et al.\ \cite{flair:morg85}. Experiment(capture): squares,
      Gieler et al.\ \cite{flair:giel91}.
      (b) Total excitation and excitation into K(4p). Theory(total
      excitation): solid curve, present results. Theory(excitation into
      K(4p)): dotted curve, 
      present results. Experiment(K(4p)): squares, Gieler et al.\
      \cite{flair:giel91}.  
      \label{fig:cs_K_l10_p} } 
  \end{center}
\end{figure}

Especially for Li atoms but also for Na a number 
of theoretical and experimental results as well as derived fits exist in
literature. Thereby these two collision systems become good candidates to
test the used method. Additionally, the results of different theoretical
approaches and 
by that their applicability can be compared. The achieved understanding of the
proton systems may be used for the discussion of the antiproton collisions
later on for which a thorough comparison is not possible due to the
sparseness of literature dealing with antiproton alkali--metal atom
collisions. The present results for proton collisions with K
complement the sparse literature data for this collision system.

\subsubsection{Ionization}
\label{sec:results_proton_ionization}
%
In figures \ref{fig:cs_Li_l10_p}, \ref{fig:cs_Na_l10_p} and
\ref{fig:cs_K_l10_p} the results of the calculations for proton collisions with
Li(2s), Na(3s) and K(4s), respectively, are presented.
The cross sections for the ionization of 
alkali metal atoms A, 
\begin{eqnarray}
  {\rm p}+{\rm A}( n_i{\rm s})& \rightarrow & {\rm A}^+\,,
\end{eqnarray}
where A stands for either Li, Na or K initially in their
ground states $n_i$s, e.g., Li(2s) are shown in the subfigures
\ref{fig:cs_Li_l10_p}a, \ref{fig:cs_Na_l10_p}a and \ref{fig:cs_K_l10_p}a,
respectively. The ionization cross section for proton collisions includes two
processes. First the ionization of the alkali metal atom due to electron
capture by the proton and second the ionization of the electron into the
continuum. The sum of both  
cross sections is sometimes also referred to as electron loss cross section.
The electron capture by the projectile is the dominant process for low
energies but vanishes fast with increasing energies and
becomes negligible for 
$E>100$\,keV. For intermediate and high energies the ionization into the
continuum is the dominant electron loss process. Therefore, in the following
discussion the present ionization calculations are also compared with electron
capture cross sections from the literature at low energies and with literature
results for ionization excluding electron capture by the proton at high
energies.


The present proton ionization cross section for Li(2s) in figure
\ref{fig:cs_Li_l10_p}a 
matches perfectly with the 2s contributions of the theoretical ionization cross
section by McCartney et al.\ \cite{flair:mcca93} and also with the
experimental results by Shah et al.\ \cite{flair:shah85}.  
The ionization cross section by Schweitzer et al.\ \cite{flair:schw99} is
somewhat smaller at high energies. 
The contribution of the 1s electrons to the ionization cross section which is
not included in the present calculations has been determined theoretically by
Sahoo et al.\ \cite{flair:saho00} and McCartney et 
al.\ \cite{flair:mcca93} and experimentally by Shah et al.\
\cite{flair:shah85}. For energies smaller than 100\,keV the contribution of
the inner shell becomes negligible compared to the one of the outer shell. For 
high energies the 1s contribution is in accordance with the difference
between the present 2s results and the Li electron ionization cross section. 
For energies smaller than 10\,keV the electron capture by the proton becomes
the dominant ionization process. Down to 4\,keV the present findings are in
good agreement with literature results shown in figure \ref{fig:cs_Li_l10_p}a
for capture and ionization. However, for energies smaller than 4\,keV the
present ionization cross section is clearly smaller than all other shown
results. This is in accordance with the 
difficulty to achieve convergence in the energy range $E \le 4$\,keV for
proton collision already discussed in section
\ref{subsec:convergence-b-dependence}. This may be a consequence of the
one-center approach which is expected to converge slowly when trying to
describe the electron capture process properly.


In figure \ref{fig:cs_Na_l10_p}a the results for the ionization of the Na atom
initially in the ground state are shown. The findings are in good agreement
with the recent results by Zapukhlyak et al.\ \cite{flair:zapu05}. Here,
especially their experimentally values match with the present curve except for
the three last data points with $E\ge 
17$\,keV which show an unexpected behavior. Their theoretical ionization cross
section agrees for energies higher than 6\,keV with the present one but is
larger for smaller energies. 
The cross section of Stary et al.\ \cite{flair:star90} which also covers
the range from low to high energies differs from the present findings as well
as from the literature results  shown here. 
The data  for electron capture by the proton shown here are consistent with
the present findings for the ionization cross section. However, the  
maximum of the fit by Tabata et al.\ \cite{flair:taba88} has a clearly higher
value. Again, it can be observed that the present ionization cross section 
for proton collisions is not fully converged for $E\le 4$\,keV leading to
smaller values in this energy range.  


In figure \ref{fig:cs_K_l10_p}a the results of the proton - K(4s)
collision calculations are presented. For potassium targets the 
literature data on proton cross sections are sparse. To the best of the
authors' knowledge no ionization cross sections for proton collisions
exists. Therefore, the present ionization cross section may be compared 
with results for electron capture. However, this is only meaningful for low
energies $ E<10$\,keV where electron capture is the dominant ionization
process. The calculations by Fritsch \cite{flair:frit84}, the experimental
data measured by Gieler et al.\ \cite{flair:giel91} as well as the fit
provided by Tabata et al.\ \cite{flair:taba88} of the electron capture cross
section are in accordance with the present ionization cross section for
$E>4$\,keV. The fitted capture cross section by Morgan et al.\
\cite{flair:morg85} results in lower values in the relevant energy range
between 4 and 10\,keV.  
In the high-energy regime the present cross section shows the same qualitative
behavior which already has been observed for Li and Na.

\subsubsection{Excitation}
\label{sec:results_proton_excitation}
%
%
In the subfigures \ref{fig:cs_Li_l10_p}b, \ref{fig:cs_Na_l10_p}b and
\ref{fig:cs_K_l10_p}b the proton excitation cross sections for Li(2s), Na(3s)
and K(4s) are shown. The total excitation of
an alkali metal atom A initially in its ground state $n_i$s, 
\begin{eqnarray}
  {\rm p}+{\rm A}(n_i{\rm s})& \rightarrow &{\rm p}+  {\rm A} (nl)\,,
\end{eqnarray}
which is the sum of all cross sections for transitions into excited
states $nl$, cf.\ Eq.\ (\ref{eq:excitation_cs}), is given. Additionally, the
cross section for the excitation process into the first excited state
$n_i$p of A,
\begin{eqnarray}
  {\rm p}+{\rm A}(n_i{\rm s})& \rightarrow &{\rm p}+  {\rm A}(n_i{\rm p})\,,
\end{eqnarray}
is given, too. The excitation into the first excited state $n_i$p is the
dominant excitation process especially at high energies. Therefore, there are
experimental data for this excitation transition. It was found in the present
investigation that it is essential in particular for high energies 
to extend the range of the impact parameter $b$ to values up to 90 
a.u.\ in order to achieve excitation cross sections which are converged with
respect to $b$.
The curves in figures \ref{fig:pb_b_lmax} and \ref{fig:pb_b_lmax_p}
for 500\,keV  already indicate that the transition probabilities for
excitation vanish slowly with increasing $b$.


The present proton excitation cross sections for Li(2s) are shown in figure
\ref{fig:cs_Li_l10_p}b.
To the best of the authors' knowledge for p - Li collisions there are no data
to compare the present total excitation cross section with. For the excitation
into Li(2p) the present results are in 
good agreement with literature data also shown in figure
\ref{fig:cs_Li_l10_p}b, except with the calculations by Stary et 
al. \cite{flair:star90}. Their findings differ for $E<6$\,keV and $E>100$\,keV
from all results shown here. The experimental data by Aumayr
\cite{flair:auma84} lie for all energies below the present calculations. On
the other hand, the calculations by Brandenburger et al.\ \cite{flair:bran98}
and also the fit provided by Wutte et 
al.\ \cite{flair:wutt97} match with the present data in the whole energy
range. Wutte et al.\ based their fit in the high-energy range on
scaled-electron-impact excitation cross sections. The calculations by Nagy et
al.\ \cite{flair:nagy00} were performed with multi-configuration wavefunctions
with an orbital basis up to $n=2$ (MC2) and up to $n=3$ (MC3).

In figure \ref{fig:cs_Na_l10_p}b p - Na(3s) collision cross sections for
the total excitation and excitation into the 3p state of the sodium atom
initially in the ground state are shown. 
The theoretical data for the total excitation cross section 
by Shingal et al.\ \cite{flair:shin87} agree well with the present results
although they show a feature around 4\,keV which is not reproduced by the
present findings. However, their excitation cross section into the Na(3p)
state follows almost 
completely the present results -- even around 4\,keV. The older calculation by
Shingal et 
al.\ \cite{flair:shin86} agrees reasonable in the energy range $4-14$\,keV but
shows a different behavior for higher and lower energies. Although the
calculations of Jain et al.\ \cite{flair:jain95} lead for all energies to
higher values their qualitative behavior is comparable to the present
results. The findings of Stary et al.\ \cite{flair:star90} show the same
behavior as their results for p - Li collisions in figure
\ref{fig:cs_Li_l10_p}b, namely, a cross 
section which is comparable around the maximum but falls off too fast for
higher and lower energies. The experimental data provided by Aumayr et al.\
\cite{flair:auma87} is in line with the present cross section for excitation
into Na(3p). It also shows a feature around 4\,keV.

%


In figure \ref{fig:cs_K_l10_p} the results of the present p - K(4s)
collision calculations are presented. 
For excitation into the K(4p) state the experimental findings of Gieler et
al.\ \cite{flair:giel91} are in good
agreement with the present calculations around the maximum but then start to
differ for $E\le4$\,keV. Their data points fall off faster while the present
result shows a behavior which has been already observed for p - Na collisions
in figure \ref{fig:cs_Na_l10_p}b. For Na the slope of the curve changes
characteristically around $E=4$\,keV. However, there is no comparable feature
for p - Li(2s) collisions. Although the excitation results -- in contrast
to ionization -- for Li and Na collision seem to be reasonable also for low
energies it is not possible to quantify how reliable the p - K(4s) excitation
cross sections for $E\le4$\,keV are. The splitting of the energy levels due to
spin-orbit coupling which is neglected in the present investigation is
supposed to be most relevant for the 4p state of K. However, the good
agreement of the present results with the experimental data by Gieler et
al.\ for $E > 4$\,keV suggests that the effect due to spin-orbit coupling
does not play a major role with respect to the level of accuracy which is
achieved by the present method.

%
%

In conclusion, the comparison of the present proton ionization and excitation
cross sections with literature data results in a good overall agreement in the
energy range $4\,{\rm keV} < E < 1000$\,keV. Thereby, the applicability of the
present method is confirmed. The findings by Stary et al.\
\cite{flair:star90}, however, differ from the present and the shown literature
results. The calculations for p - K collisions complement the data provided
by the sparse literature on this collision system.

\subsection{Cross sections for antiproton collisions }

\begin{figure}[t]
  \begin{center}
    \includegraphics[width=0.49\textwidth,angle=0]{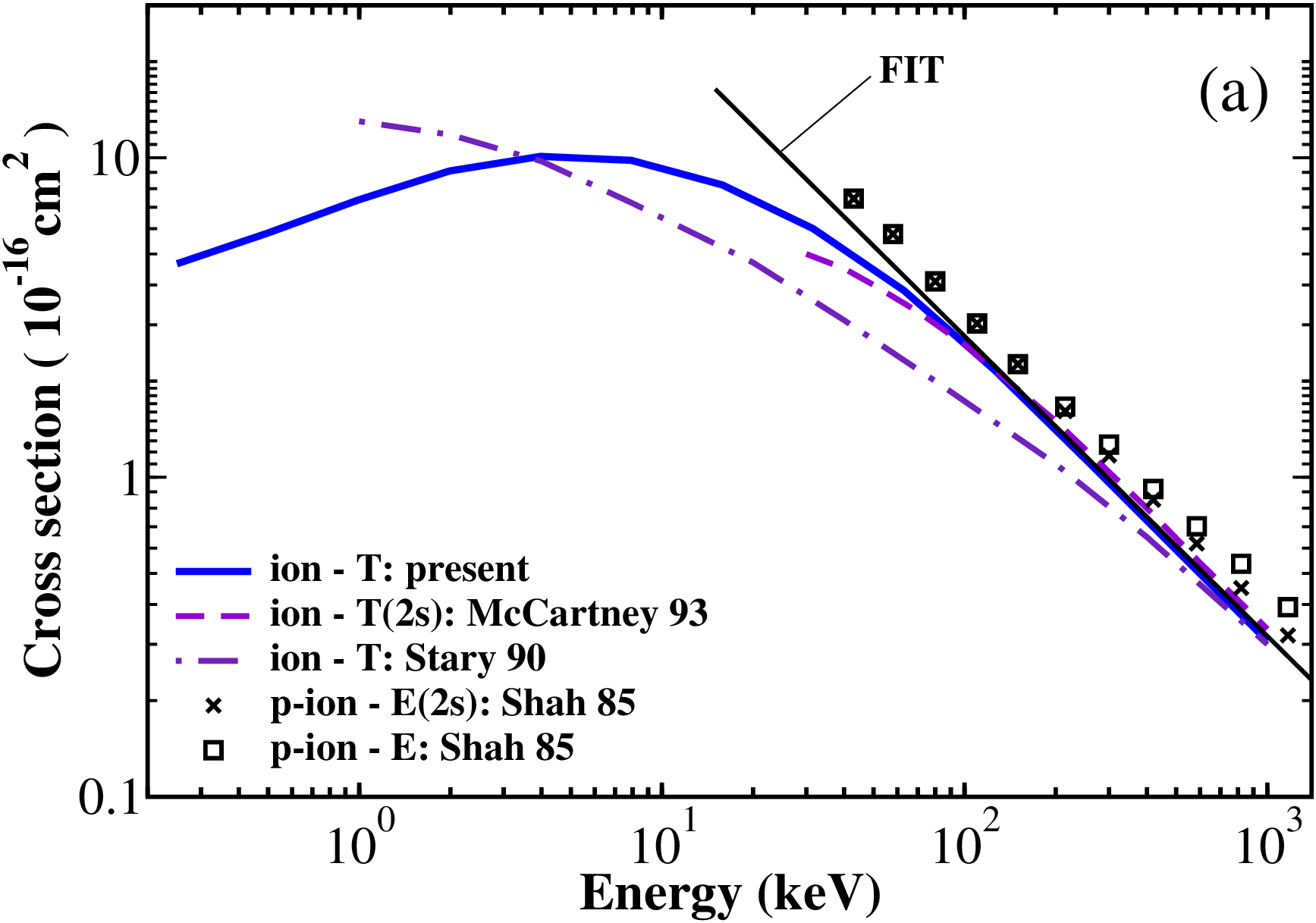} 
    \includegraphics[width=0.49\textwidth,angle=0]{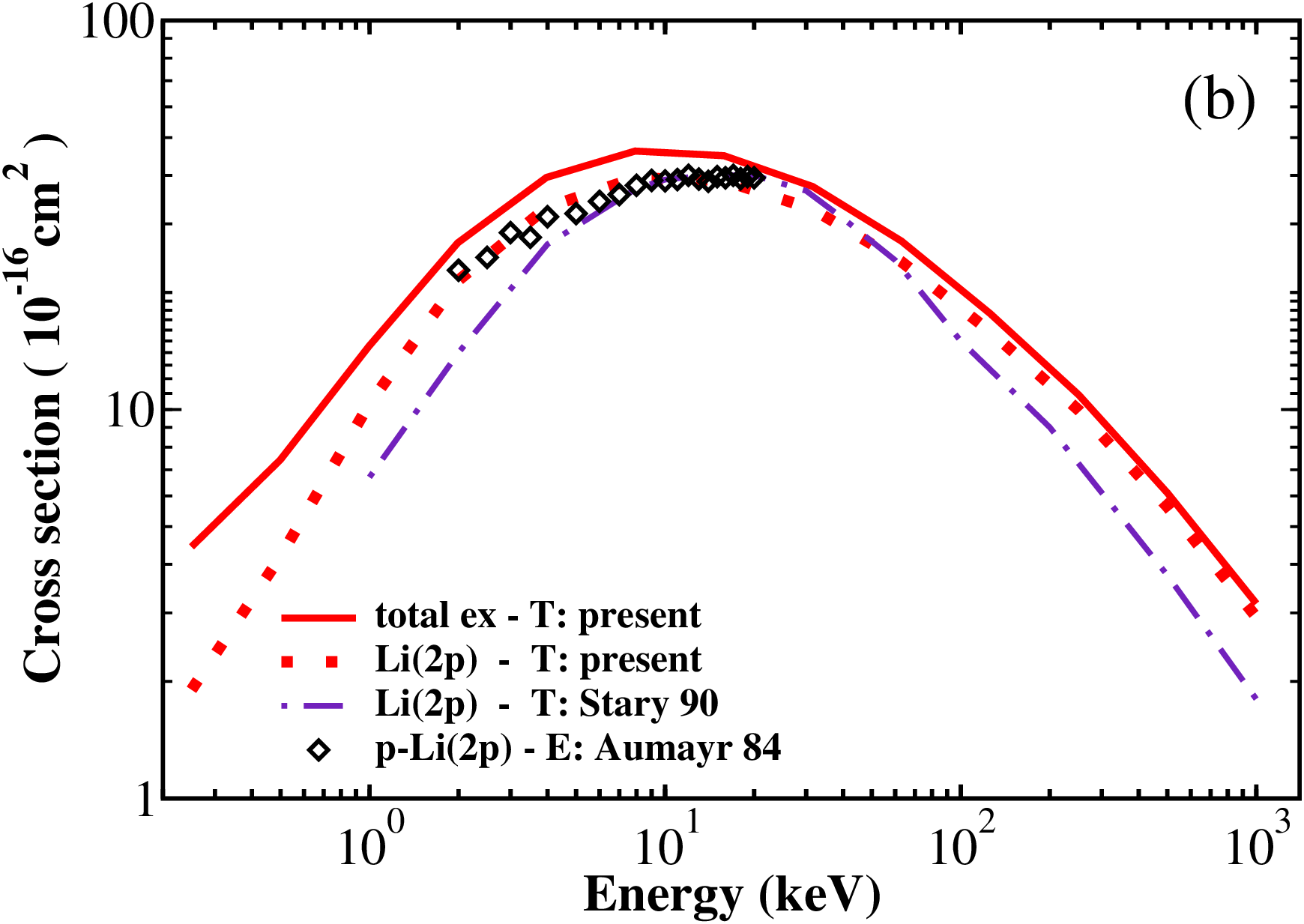} 
    \caption{(Color online) \=p - Li(2s): (a) Ionization. Theory: solid
      curve, present results; dashed dotted curve, Stary et al.\
      \cite{flair:star90}; dashed curve, McCartney et al.\
      \cite{flair:mcca93}. Experiment: crosses (p-Li(2s)), Shah et
      al.\  \cite{flair:shah85}; squares (p-Li(2s) and p-Li(1s)), Shah et al.\ 
      \cite{flair:shah85}.  (b) Total excitation and excitation into
      Li(2p). Theory(total excitation): solid curve, present
      results. Theory(excitation into Li(2p)): dotted curve, present results;
      dashed dotted curve, Stary et al.\ \cite{flair:star90}.
      Experiment(excitation into Li(2p)): diamonds (p-Li(2s)), Aumayr et al.\
      \cite{flair:auma84}.  
      \label{fig:cs_Li_l8} } 
  \end{center}
\end{figure}
\begin{figure}[t]
  \begin{center}
    \includegraphics[width=0.49\textwidth]{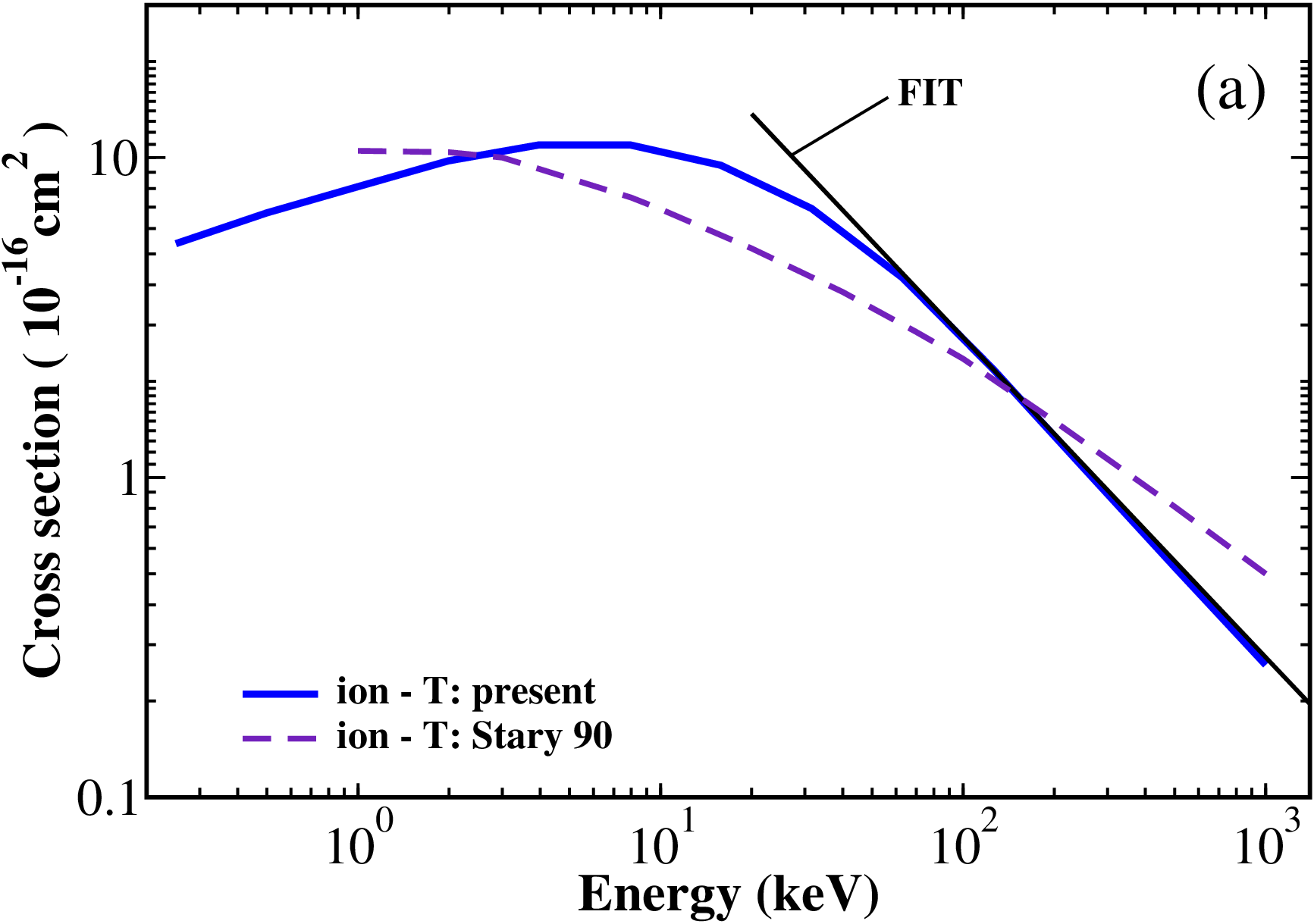}
    \includegraphics[width=0.49\textwidth]{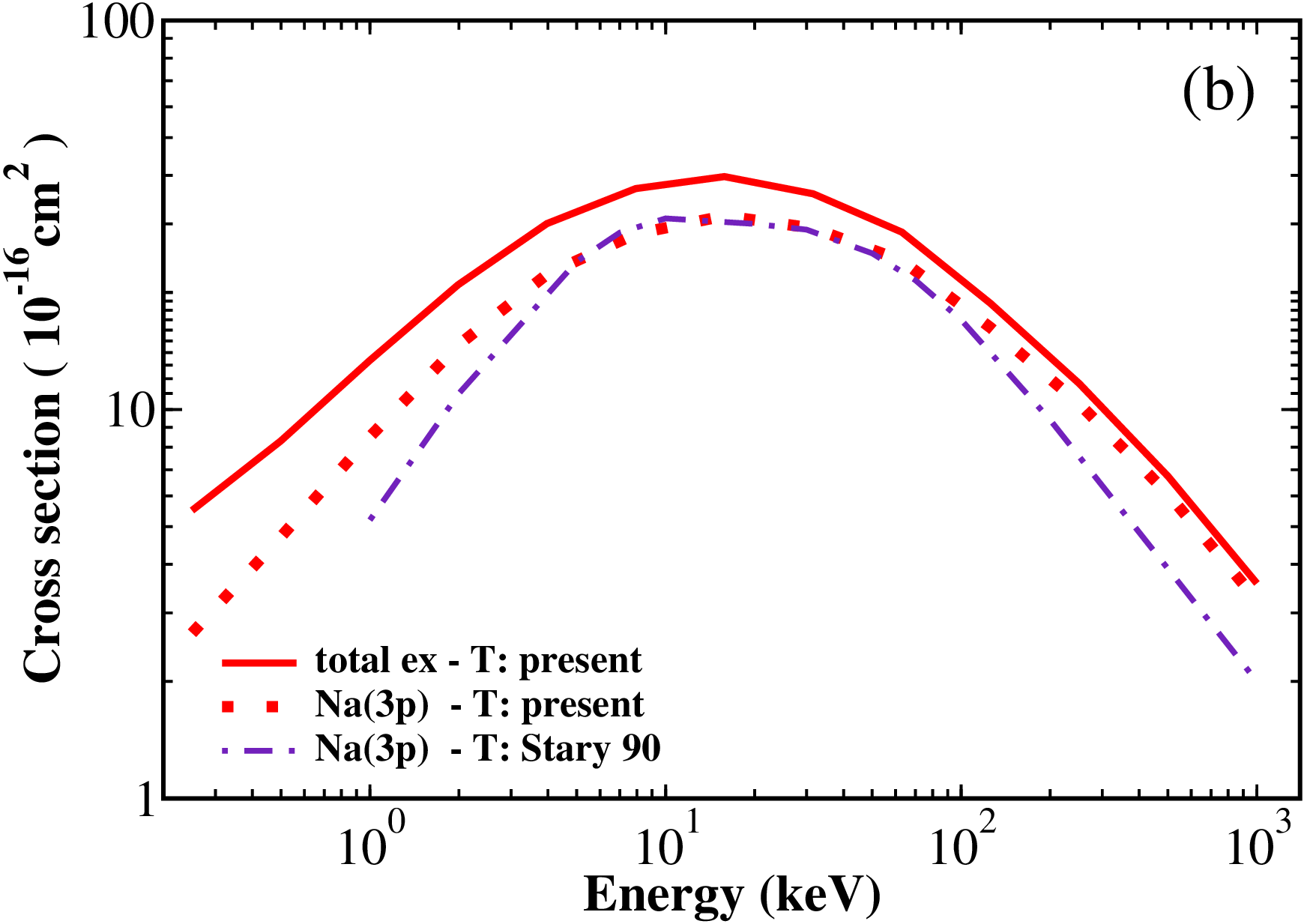}
    \caption{(Color online) \=p - Na(3s) (a) Ionization. Theory: solid
      curve, present results; dashed curve, Stary et al.\
      \cite{flair:star90}.  (b) Total excitation and excitation into
      Na(3p). Theory(total excitation): solid curve, present
      results. Theory(excitation into Na(3p)): dotted curve, present results.
      \label{fig:cs_Na_l8} }  
  \end{center}
\end{figure}
\begin{figure}[t]
  \begin{center}
    \includegraphics[width=0.49\textwidth]{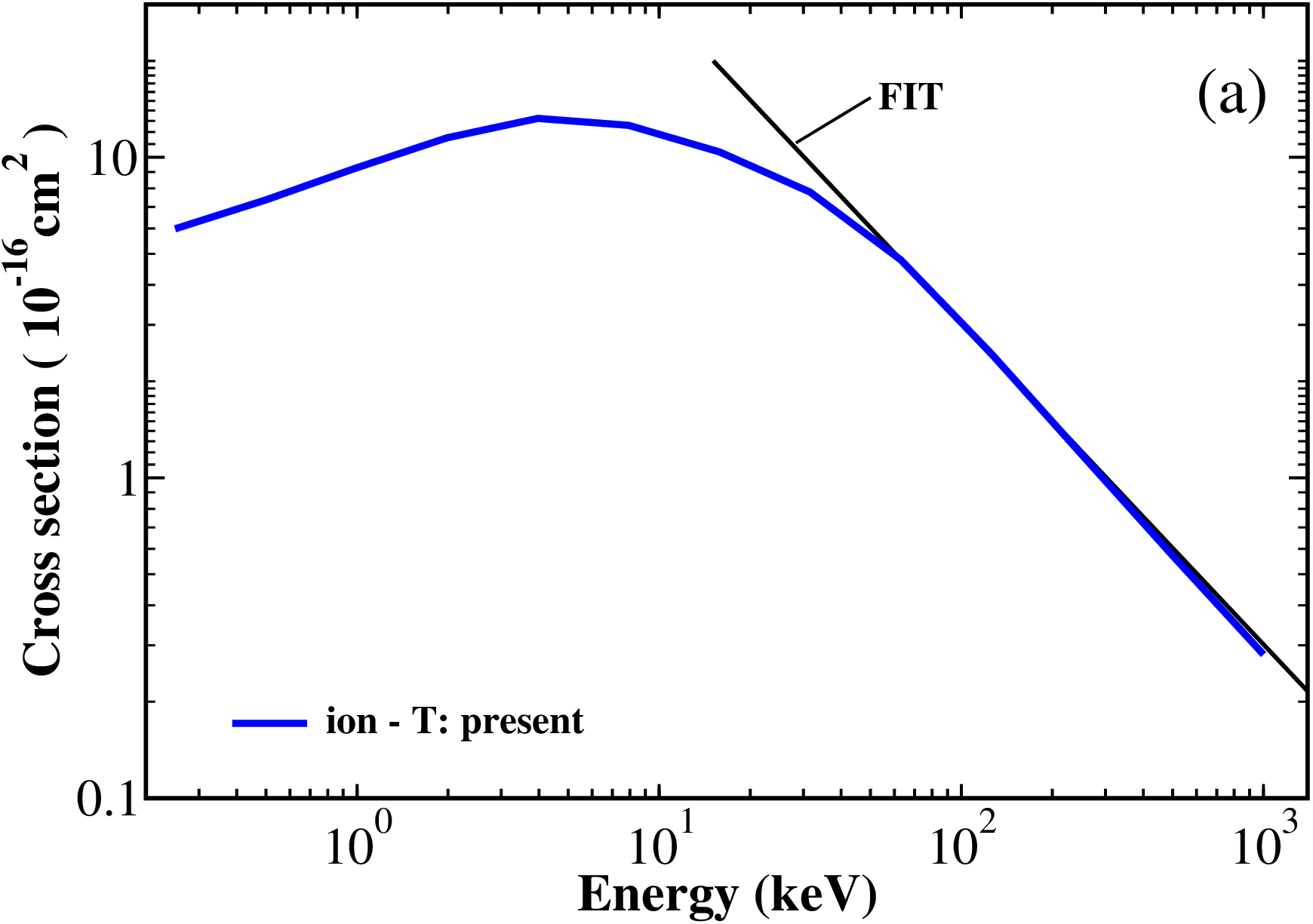}
    \includegraphics[width=0.49\textwidth]{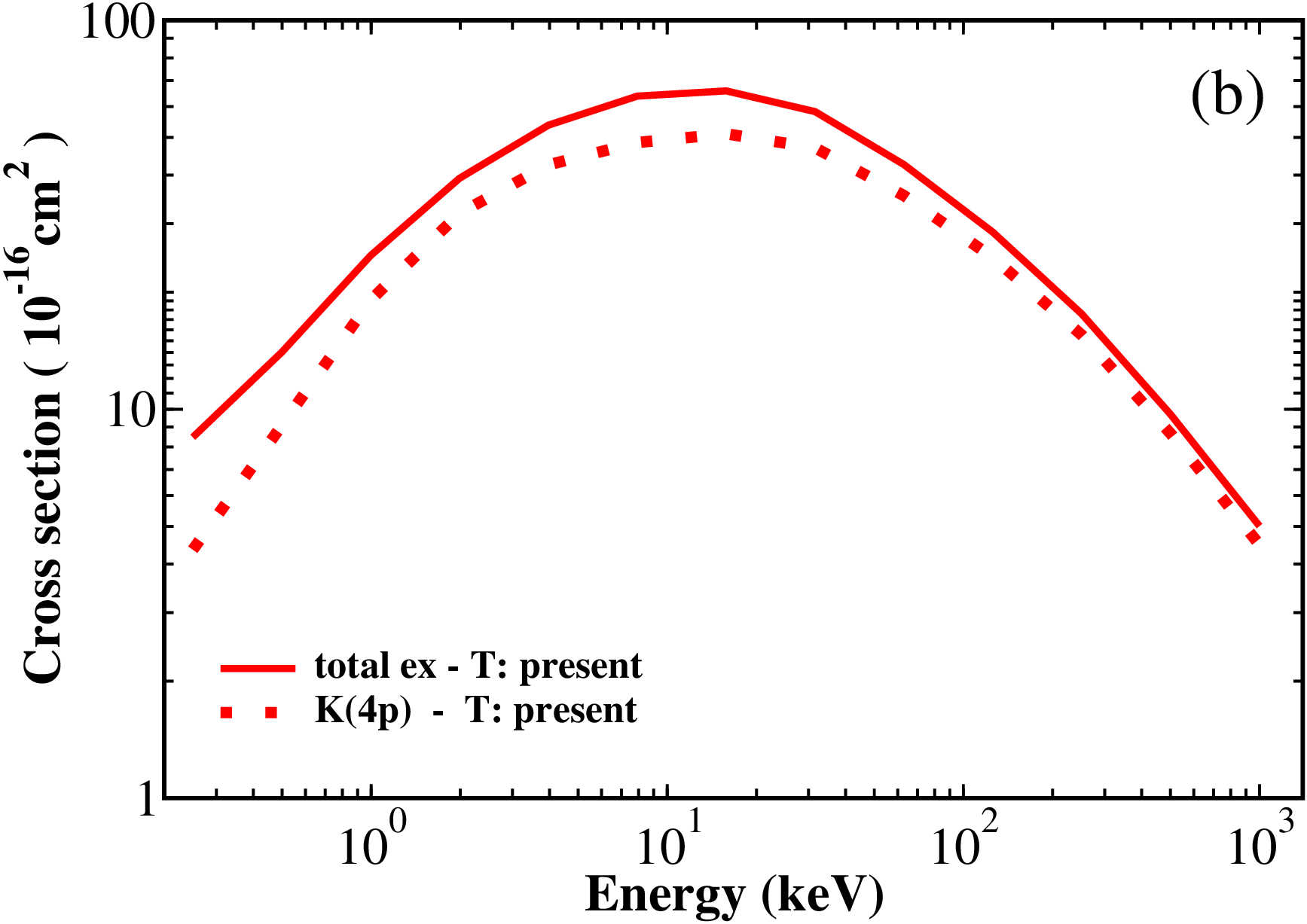}
    \caption{(Color online) \=p - K(4s) (a) Ionization. Theory: solid
      curve, present results.  (b) Total excitation and excitation into
      K(4p). Theory(total excitation): solid curve, present
      results. Theory(excitation into K(4p)): dotted curve, present results.
      \label{fig:cs_K_l8} } 
  \end{center}
\end{figure}

Only very few data for antiproton--alkali metal atom collisions exist in the
literature. Cross sections are available for the ionization of Li(2s) and
Na(3s)  as well as for excitation into Li(2p) and Na(3p) by Stary et al.\
\cite{flair:star90}. Furthermore, there are ionization cross sections for \=p
- Li collisions calculated by McCartney et al.\ \cite{flair:mcca93}. 
However, no cross section exists for ionization or excitation into K(4p) for K
targets and also for all three considered target atoms there are no 
total excitation cross sections to which the present results could be compared.

\subsubsection{Ionization and excitation}
\label{sec:results_antiproton_ionization}
%

%
The ionization cross sections for antiproton collisions with the target atoms
Li(2s), Na(3s) and K(4s) are shown in figures \ref{fig:cs_Li_l8}a,
\ref{fig:cs_Na_l8}a and \ref{fig:cs_K_l8}a, respectively, and for excitation
accordingly in figures \ref{fig:cs_Li_l8}b, \ref{fig:cs_Na_l8}b and
\ref{fig:cs_K_l8}b. 
%
%
The theoretical results for ionization in \=p - Li collisions by McCartney et 
al.\ \cite{flair:mcca93} agree well with the present findings. However, they
only cover the high-energy regime $E>30\,$keV.
The calculated antiproton ionization cross sections for Li and Na targets by
Stary et al.\ \cite{flair:star90} both differ considerably from the present
findings. Their \=p - Li ionization cross 
section behaves differently for low to intermediate energies but seems to
converge to the present findings for high energies. Their \=p - Na ionization
cross section, however, shows a different behavior compared to the present
curve in the whole energy range.


The cross sections for the excitation into the first excited state for Li and
Na target atoms calculated by Stary et al.\ \cite{flair:star90} both share the
same features. Their cross sections agree with the present curves around the
maxima at $E\approx 10$\,keV and $E\approx 15$\,keV for Li and Na, respectively,
but fall off faster for lower and higher energies. The same behavior has been
observed in the case of proton collisions 
in figures \ref{fig:cs_Li_l10_p}b and  \ref{fig:cs_Na_l10_p}b for excitations 
into Li(2p) and Na(3p), respectively. Therefore, their results differ once
more from the outcome of the present investigation. 
The aim of Stary et al.\ was to obtain results comparable to literature data
but using smaller basis sets within an optical potential approach adapted to
this problem. A Feshbach 
projector formalism for the solution of the time-dependent Schr\"odinger 
equation leading to a finite set of coupled-channel equations with complex
potentials was used. Thereby, two conditions were assumed to be
fulfilled. First, the interactions occur instantaneously and second, the
energy distribution of the Q-space which is the complement of the finite model
space has a peak leading to the assumption of an average Q-space energy $\bar
\epsilon$. Furthermore, a scaling factor is used which restores the correct
energy dependence of the optical potential and which is determined at high
impact energies. 
Since the present results for the proton case seem to be more
reliable than their one-center calculations the present results for antiproton
collisions with Li and Na are also considered to be more reliable than
theirs. If their solutions are converged as it was claimed by Stary et al.\
then either not both of the above mentioned conditions are  fulfilled or the
introduced scaling factor has a different functional behavior.


To the best of the authors' knowledge no literature on \=p - K cross
sections exist for the considered energy range. The present cross sections for
excitation and ionization of K in figure \ref{fig:cs_K_l8} show a
qualitatively similar behavior like for \=p - Na collisions in figure
\ref{fig:cs_Na_l8} but with higher values throughout the energy range. 


Until now, experimental results for the antiproton--alkali metal atom collision
systems are completely missing in the considered energy range. It is
remarkable that the experimental data of Aumayr \cite{flair:auma84} for
excitation into Li(2p) by proton collisions fits better to the present
antiproton than proton data.

\subsubsection{Comparison of antiproton with proton cross sections }
\label{sec:results_comparison_antiproton_proton}

%
%

%
%

\begin{figure}[t]
  \begin{center}
    \includegraphics[width=0.49\textwidth]{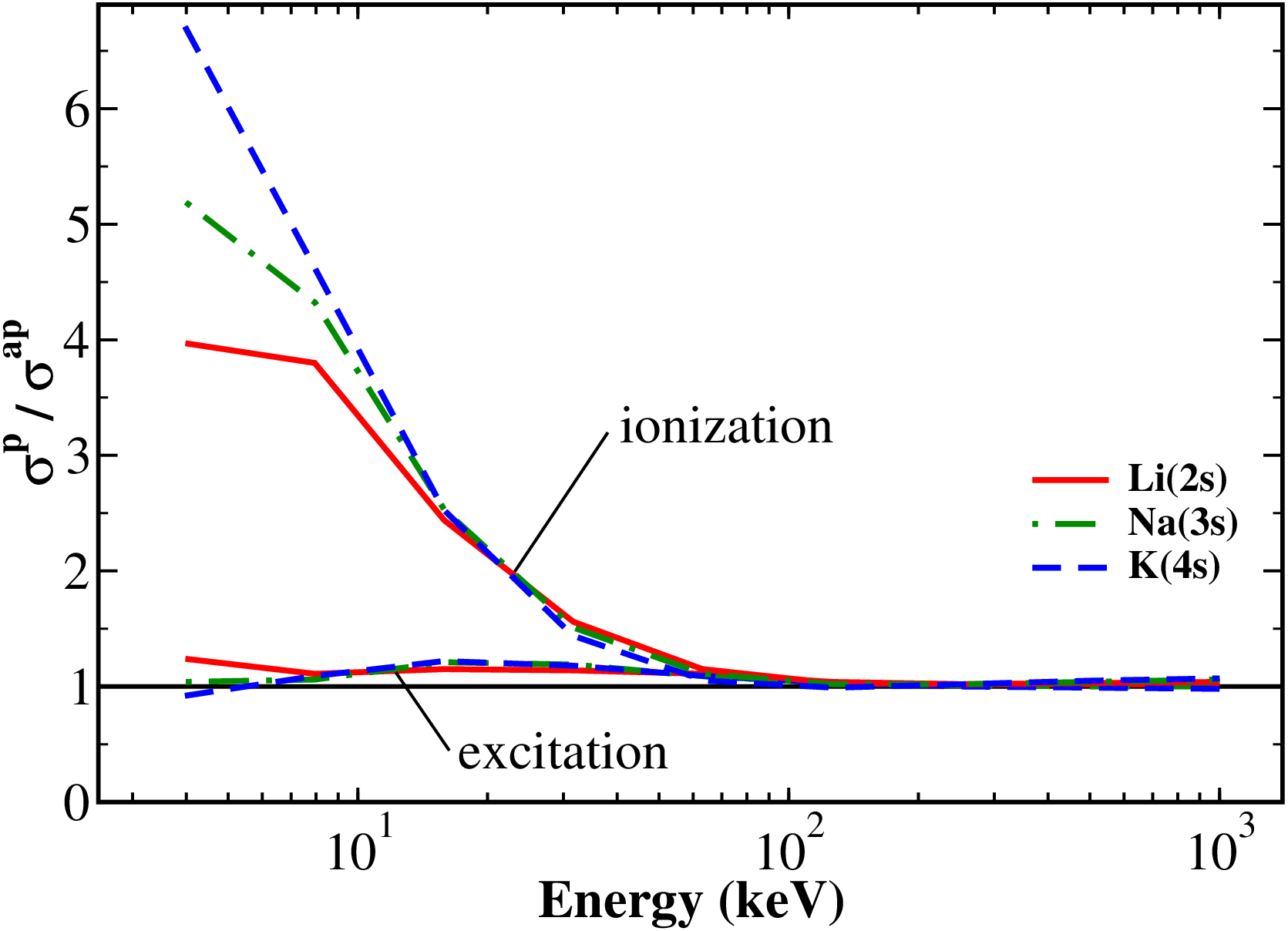}
    \caption{(Color online) Ratio of proton to antiproton cross sections
      $\sigma^{\rm p}\, /\, \sigma^{\rm ap} $. The ratios of the ionization and
      excitation cross sections for the target atoms Li(2s), Na(3s) and K(4s)
      are given. 
      \label{fig:cs_CMP_ap_p} } 
  \end{center}
\end{figure}

While for sufficiently high energies a similar behavior for proton and
antiproton cross sections is to be expected the collision processes should
differ for lower energies. In contrary to the proton collisions no
electron capture by the projectile is possible for antiprotons. Since the
electron capture is the dominant ionization channel for low energy proton
collisions  noticeable differences especially for the antiproton ionization
cross sections can be expected in the low-energy regime. In what follows the
antiproton and proton cross sections are compared in some detail for high,
intermediate and low impact energies. In figure \ref{fig:cs_CMP_ap_p} the
ratios of proton to antiproton cross sections for ionization and excitation is
given for the three considered target atoms. 

To begin with the comparison focuses on the high energy behavior of the
antiproton and proton cross sections. In the validity range of the
first Born approximation no differences in the cross sections for
different projectiles like electrons, protons and
antiprotons with the same velocity are expected because in this approximation
the cross sections only depend on the absolute value of the projectile
charge. It is a high energy approximation.
%
%
%
\begin{table}[b]
  \centering
  \begin{tabular}{c@{\hspace{0.2cm}}|@{\hspace{0.2cm}}cc@{\hspace{0.2cm}}c}
    \hline
    \hline
    Atom & $E_0$&\ $\sigma_{\rm ion}(E_0)$\ &a\\
    \hline
    Li(2s) &141.3 &2  &-0.9386 \\
    Na(3s) &138   &2  &-1      \\
    K (4s) &151   &2  &-1      \\
    \hline
    \hline
  \end{tabular}
  \caption{Parameters for the description of the ionization cross section
    for the energy range $100\,{\rm keV} < E_0 <1000\,{\rm keV}$ using the
    fit formula (\ref{eq:sigma_ion_fit}). $a$ is a dimensionless fit
    parameter. The energy $E_0$ is given in keV and the cross section
    $\sigma_{\rm ion}(E_0)$ for $E_0$ in units of $10^{-16}\, {\rm
      cm}^2$. 
    \label{tb:fit-parameter}} 
\end{table}
A linear decrease of the ionization cross section for all three alkali
metal atoms can be observed on a doubly-logarithmic 
scale for high energies $100\,{\rm keV} < E < 1000\,{\rm keV}$ for protons as
well as for antiprotons. Therefore, the general fit formula
\begin{equation}
  \label{eq:sigma_ion_fit}
  \sigma_{\rm ion}(E) = \sigma_{\rm ion}(E_0)\,\left(\frac{E}{E_0}\right)^a
\end{equation}
for ionization cross sections in this energy range can be proposed,   
where $\sigma_{\rm ion}(E_0)$ is the ionization cross section for an arbitrary
$E_0$ in the range $100\,{\rm keV} < E_0 <1000\,{\rm keV}$ and $a$ is a fit
parameter which gives the slope of the linear curve on a doubly-logarithmic
scale. The fit parameters which may be proposed for the three alkali metals
colliding with protons are given in table \ref{tb:fit-parameter}. The fits for
Na and K reveal a direct proportionality between the ionization cross section
and the inverse of the energy,
\begin{equation}
   \sigma_{\rm ion}(E)=\frac{\sigma_{\rm ion}(E_0)\,E_0}{E} \propto
   \frac{1}{E} \, ,
\end{equation}
in the considered high energy regime. This proportionality holds 
approximately also for the present Li ionization cross section.
The proposed fits which are also shown in figures \ref{fig:cs_Li_l10_p}a,
\ref{fig:cs_Na_l10_p}a and \ref{fig:cs_K_l10_p}a match well with the
calculated ionization cross sections for $E\ge150$\,keV. 
These fits obtained for the proton case are also shown in  figures
\ref{fig:cs_Li_l8}a, \ref{fig:cs_Na_l8}a and \ref{fig:cs_K_l8}a in order to
compare them with the antiproton ionization cross sections. It can be seen
that the proton fits match remarkably well with the antiproton ionization
results for energies higher than 150\,keV. Therefore, the antiproton
ionization cross sections also decrease proportional to $E^{-1}$ where this
proportionality  again holds only approximately for Li targets.  This means
that for energies higher than 150\,keV  
no specific features are expected for antiproton ionization cross
sections with the considered alkali metal atoms. And in turn for these
energies the treatment of proton collisions should be sufficient which is
especially in the case of experimental studies less demanding. 

However, both systems differ strongly regarding ionization for energies lower
than 100\,keV due to 
the electron capture process which is only possible for protons. The proton
cross section is strongly enhanced which can also be
seen in figure \ref{fig:cs_CMP_ap_p}. The maxima of the proton and antiproton
ionization cross sections approximately at 45 and 10 [in $10^{-16}\,{\rm
  cm}^{2}$], respectively,  for Li targets differ by a factor 4.5. The
ionization maxima for Na and K targets differ approximately by a factor 5.5
and 6.5, respectively. The maxima are all located between 4 and 6\,keV where
the proton maxima tend to occur at lower energies than the corresponding
antiproton maxima.



%
%
A comparison of the present excitation cross sections for proton and
antiproton  collisions yields that they also agree for high energies
$E>150$\,keV. The antiproton maximum for Li targets lies around 10\,keV and is
10\% lower than for proton collisions. The antiproton maxima for Na and K are
situated at approximately 15\,keV with $\approx 20\%$ smaller values than for
the proton case.  
But below their maxima the Na and K excitation curves for 
antiprotons and protons  excitation cross section have comparable values.

The most striking feature of figure \ref{fig:cs_CMP_ap_p} is that the ratios of
the proton to antiproton ionization cross sections increase strongly for
low-energy collisions while the ratios for excitation only vary comparably
weakly around 1. In the case of ionization the electron capture channel becomes
important for low-energy proton collisions leading to large ionization cross
sections compared to antiproton collisions. In the case of excitation for both
projectiles the same channels are open.

%
%
It can be concluded for the antiproton cross sections that the present results
complement and improve the existing data on antiproton--alkali metal atom
collisions. While the excitation cross sections are comparable for proton and
antiproton projectiles the proton ionization cross sections are strongly
enhanced at low energies due to electron capture. For high energies
$E>150$\,keV proton and antiproton collisions with Li, Na and K result in the
same ionization cross sections which decrease proportional to $E^{-1}$.

\subsection{Comparison of the antiproton cross sections }
\begin{figure}[t]
  \begin{center}
    \includegraphics[width=0.49\textwidth]{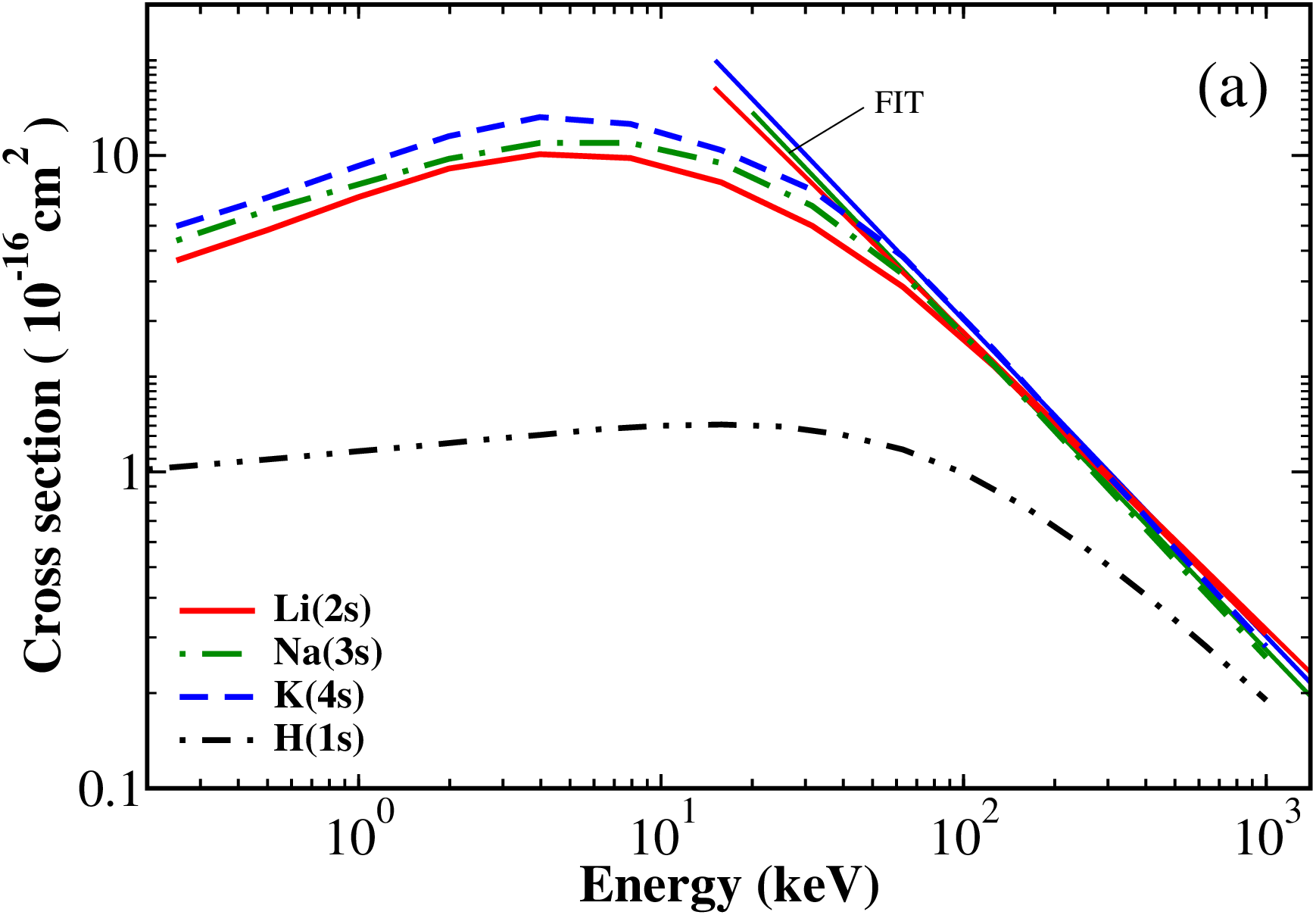}
    \includegraphics[width=0.49\textwidth]{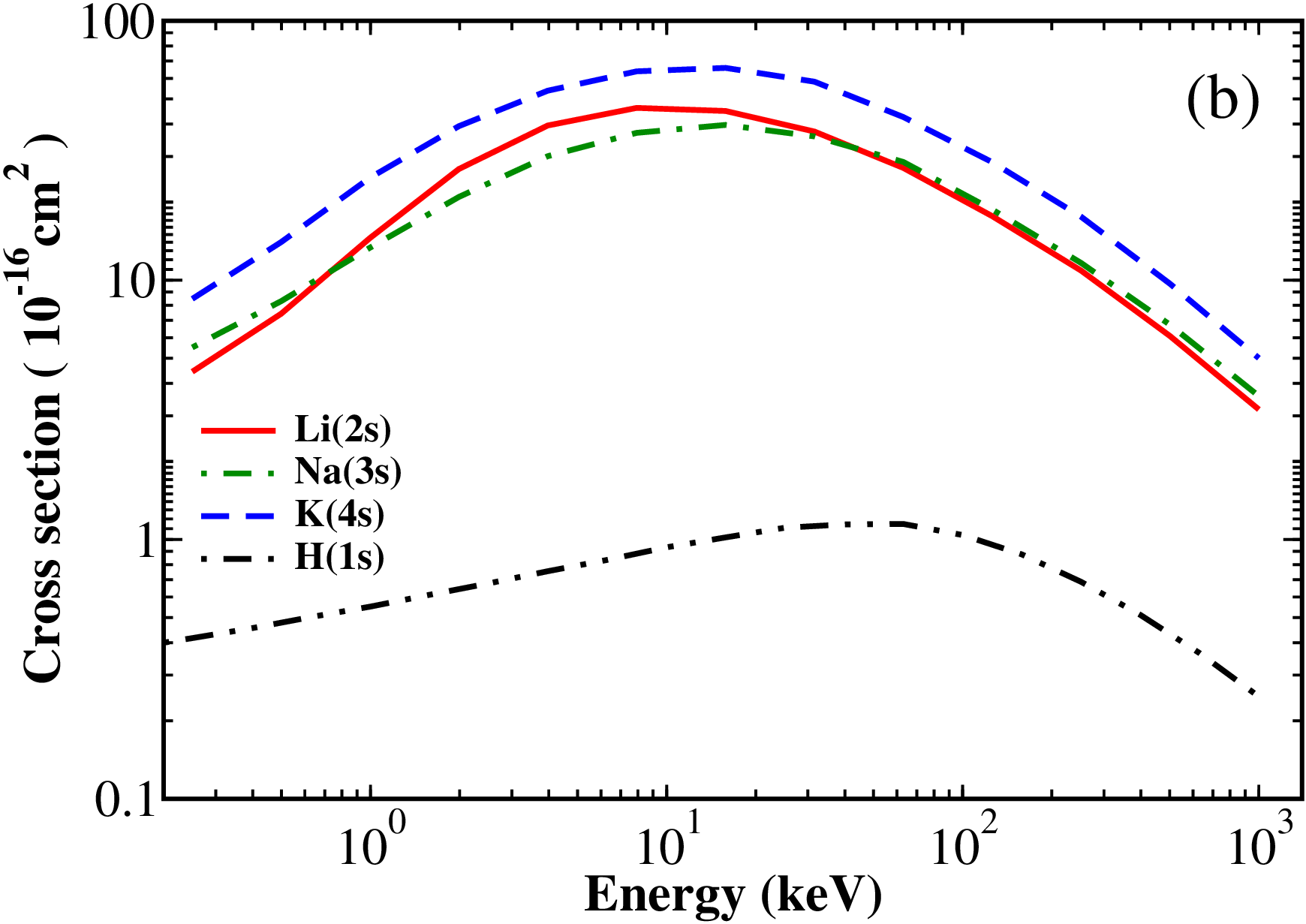}
    \caption{(Color online)  (a) Single-electron ionization cross section for
      antiprotons colliding with Li(2s), solid curve, Na(3s), dashed dotted
      curve, K(4s), dashed curve, and H(1s), dashed doubly-dotted
      curve. Additionally, the fits describing the high-energy behavior of
      proton scattering are given.  
      (b) Total excitation cross section for antiprotons colliding with
      Li(2s), solid curve, Na(3s), dashed dotted curve, K(4s), dashed
      curve, and H(1s), dashed doubly-dotted curve.  
      \label{fig:cs_Cmp_l8} } 
  \end{center}
\end{figure}

In figure \ref{fig:cs_Cmp_l8}a the ionization cross sections for the three
alkali metal atoms Li, Na and K colliding with antiprotons are plotted
together with the high-energy 
fits extracted earlier from the proton calculations. The qualitative behavior
of the cross sections for these atoms is similar in the whole energy range.
All curves converge to their corresponding proton results for energies
$E\ge150$\,keV which lie close to each other. The differences between
the heights of the ionization cross sections for $E<100$\,keV as well as the
ordering of the curves may be explained by the different ionization energies:
${\rm Li} = 0.198$, ${\rm   Na} = 0.189$, ${\rm K} = 0.160$ which can also be
found in table  
\ref{tb:binding-energies}. All maxima lie around 4 to 6\,keV which is
somewhat below the average velocity of the valence electrons and have
far lower values compared to the proton collision systems.

%

In figure \ref{fig:cs_Cmp_l8}b the total excitation cross sections for the
three alkali metals are compared. Although the overall behavior of their
excitation cross sections are similar they differ in detail. 
On an absolute scale the values of the antiproton excitation maxima for the
three atoms differ considerably [in $10^{-16}$\,cm$^2$]: ${\rm Li} = 46.2$,
${\rm Na} = 39.7 $, ${\rm K} = 65.9 $, i.e., the maximum for K is 66\% higher
than that for Na. This can be made plausible by comparing
the energy differences of the ground states to the first excited states: ${\rm
  Li} = 0.068$, ${\rm Na} = 0.089$, ${\rm   K} = 0.059$ since 
the $n_i{\rm s} \rightarrow n_i{\rm p}$ transition is the dominant
excitation channel. 

%
The cross sections for ionization and excitation of
the hydrogen atom by antiproton impact were also calculated which are
presented in figure \ref{fig:cs_Cmp_l8}, too. The qualitative behavior of the 
hydrogen cross sections is comparable to those of the alkali metal atoms
reflecting the shell structure of the alkali metal atoms with an outer valence
electron in an s state. However, the absolute values of the cross sections
differ clearly. First, the cross sections for hydrogen are much smaller due
to the tighter binding of the electron which leads to a higher ionization
energy and a smaller spacial extension. Second, the maxima are shifted to
higher impact energies which can be 
explained by the higher average velocity of the electron in the ground state
of hydrogen. For high energies ($E\geq 1000$\,keV) the
ionization cross section of hydrogen seem to approach those of the alkali
metal atoms. At these energies the ionization cross section of hydrogen is
expected to decrease like $E^{-1}\log E$.

%
%

%

\section{Conclusion}
\label{sec:summary}
Time-dependent close-coupling calculations of ionization and excitation
cross sections for antiproton and proton collisions with the alkali metal atoms
Li(2s), Na(3s) and K(4s) have been performed in a wide energy range from 0.25 to
1000\,keV. The 
target atoms are treated as effective one-electron atoms using a model
potential. The total wave function is expanded in an one-center approach in
eigenfunctions of the 
one-electron model Hamiltonian of the target atom. The radial part of the
basis functions is expanded in B-spline functions and the angular part in
a symmetry-adapted sum of spherical harmonics. The collision process is
described in the classical trajectory approximation. In the present calculations
the results converged faster for  collisions involving antiprotons than
protons, faster for high than for low energies and faster for excitation
than for ionization. Good agreement with literature data has been achieved for
the proton--alkali metal atom cross sections for $E\ge4$\,keV. However, for 
antiproton--alkali metal atom collisions literature data are sparse. The
comparison to the 
calculations of antiproton collisions with Li and Na by Stary et al.\ shows
the same disagreement with the present findings as it was found for their
proton collision results. In
view of this disagreement with literature data for proton collisions it can
be stated that the calculations by Stary et al.\ were either not fully
converged or the assumed conditions not fulfilled. To 
the best of the authors' knowledge the first cross sections for \=p - K 
collisions in the considered energy range are presented. The ionization cross
sections for protons and antiprotons differ considerably for energies smaller
than 100\,keV due to the electron capture process which is only possible for
protons and is the dominant ionization channel at low energies. The qualitative 
behavior of the antiproton cross sections is comparable for all three alkali
metal atoms but differs in the absolute values depending on the atom-specific
ionization and excitation energies. A comparison with hydrogen as target atom
yields the same characteristics as for the alkali metals due to the common s
state structure. However, the cross sections of hydrogen have much lower
values and the hydrogen ionization and excitation maxima are shifted 
to higher impact energies because of the tightly bound 1s electron.
For the proton ionization cross sections a simple fit formula is proposed for
the energy range from 150 to 1000\,keV which also describes the properties of
the antiproton ionization cross sections in this energy range well. The fit
reveals that the ionization cross sections decrease proportional to $E^{-1}$
in this energy range.

\begin{center}
 {\bf ACKNOWLEDGMENTS} 
\end{center}
The authors are grateful to BMBF (FLAIR Horizon) and {\it Stifterverband f\"ur
  die deutsche Wissenschaft} for financial support.



\end{document}